\documentclass[aps,pre,superscriptaddress,twocolumn]{revtex4-1}
\usepackage{amsmath}
\usepackage{xcolor}
\usepackage{graphicx}
\usepackage{dcolumn}
\usepackage{bm}
\usepackage{hyperref}

\usepackage{amsfonts}
\usepackage{multirow}
\usepackage{color}
\usepackage{times}
\usepackage{caption}
\usepackage{subcaption}

\begin{document}

\title{Controlling the spread of deception-based cyber-threats on time-varying networks}

\author{Nicol\`o Gozzi}
\affiliation{ISI Foundation, Turin, Italy}

\author{Nicola Perra}
\email[]{n.perra@qmul.ac.uk}
\affiliation{School of Mathematical Science, Queen Mary University of London, London, UK}
\affiliation{The Alan Turing Institute, London, UK}

\date{\today}

\begin{abstract}
We study the efficacy of strategies aimed at controlling the spread of deception-based cyber-threats unfolding on online social networks. We model directed and temporal interactions between users using a family of activity-driven networks featuring tunable homophily levels among gullibility classes. We simulate the spreading of cyber-threats using classic Susceptible-Infected-Susceptible (SIS) models. We explore and quantify the effectiveness of four control strategies. Akin to vaccination campaigns with a limited budget, each strategy selects a fraction of nodes with the aim to increase their awareness and provide protection from cyber-threats. The first strategy picks nodes randomly. The second assumes global knowledge of the system selecting nodes based on their activity. The third picks nodes via egocentric sampling. The fourth selects nodes based on the outcome of standard security awareness tests, customarily used by institutions to probe, estimate, and raise the awareness of their workforce. We quantify the impact of each strategy by deriving analytically how they affect the spreading threshold. Analytical expressions are validated via large-scale numerical simulations. Interestingly, we find that targeted strategies, focusing on key features of the population such as the activity, are extremely effective. Egocentric sampling strategies, though not as effective, emerge as clear second best despite not assuming any knowledge about the system. Interestingly, we find that networks characterized by highly homophilic interactions linked to gullibility might expand the range of transmissibility parameters that allows for macroscopic outbreaks. At the same time, they reduce the reach of these spreading events. Hence, rather isolated patches of the network formed by highly gullible individuals might provide fertile grounds for the propagation and survival of cyber-threats.
\end{abstract}

\maketitle

Deception-based attacks such as phishing, baiting, and file masquerading have become one of the most diffuse types of cyber-threats~\cite{GTREPORT24,kayes2017privacy, gupta2017fighting,heartfield2016predicting,heartfield2016you,zimmermann2019moving,heartfield2016taxonomy,heartfield2018protection}. These are designed around ingenious strategies that target human nature. For example, the classic phishing scheme tries, via trusted access, to lure victims to open a malicious link and/or to download a malware which might affect personal accounts and devices, reveal sensitive data, and, unbeknownst to the victims, allow the threat to spread further. Worryingly, the recent advancements in generative artificial intelligence are offering new unprecedented opportunities to malicious actors to enhance and scale-up their criminal activities~\cite{Falade2023Decoding,grbic2023social}.

The extant literature devoted at modeling these processes and gathering insights to contrast them is vast, but presents two main limitations. First, most of previous work neglects the temporal nature of online contact patterns focusing instead on aggregated networks~\cite{holme11-1,Holme2015ModernTN,newman2002email,balthrop2004technological,cohen03-1}. However, the order and concurrency of interactions are key factors shaping the characteristics of a wide range of spreading process on networks. Neglecting time-varying patterns in favor of static (i.e., aggregated) representations may lead to an overestimation of the spreading potential of such threats~\cite{barrat2015face,perra12-2,perra12-1,ribeiro12-2,liu2014controlling,PhysRevE.87.032805,10.1063,starnini13-1,starnini_rw_temp_nets,valdano2015analytical,scholtes2014causality,Williams160196,rocha2014random,takaguchi2012importance,rocha2013bursts,ghoshal2006attractiveness,sun2015contrasting,mistry2015committed,pfitzner13-1,takaguchi12-1,takaguchi2013bursty,holme2014birth,holme2015basic,wang2016statistical,gonccalves2015social}. Despite this general trend, it is important to mention a few exceptions. Ref.~\cite{Brett_2019} modeled the spreading of computer viruses on time-varying networks featuring homophily. Ref.~\cite{wang2009understanding} studied the spreading of computer viruses via temporal Bluetooth connections. Ref.~\cite{prakash10-1} explored the propagation of computer viruses in general time-varying networks. 

As a second limitation, most of the literature assumes users to be equally susceptible (i.e., gullible). However, recent empirical studies showed that susceptibility to cyber-threats is not homogeneous and may depend on factors such as age, digital proficiency, or familiarity with online social networks, among others~\cite{heartfield2016you, heartfield2017eye}. Also in this case we find a few exceptions in the literature, such as Ref.~\cite{Brett_2019} and Ref.~\cite{peng2017immunization}. 

Here, we build on the theoretical framework developed in Ref.~\cite{Brett_2019} and expand it to investigate the effectiveness of different strategies devoted to control the spreading of deception-based attacks. To this end, we imagine a large corporation or institution facing the challenge of protecting their digital infrastructure against cyber-threats. Following Ref.~\cite{Brett_2019}, we model the temporal interactions between users in the corporation adopting a family of activity-driven networks~\cite{perra12-1,karsai13-1,ubaldi2015asymptotic,tizzani2018epidemic}. Nodes are characterized by an activity (capturing their propensity to initiate communications), and by a membership to a gullibility class (which affects both the probability of falling for a deception-based attack and the rate of recovery, if affected). As mentioned, susceptibility is linked to several users' features. In our hypothetical scenario, categories might be linked to the organizational structure of the corporation. We can imagine that people tend to interact more with others in the same department and that departments tend to host individuals with similar computer proficiency. For example, IT departments are typically formed by professionals who are more aware of cyber-security than the average. To account for these aspects, the model includes contacts' homophily: the membership to a category influences the link creation process~\cite{mcpherson2001birds,Brett_2019}. We model the potential spreading of deception-based cyber-threats within the corporation considering a classic Susceptible-Infected-Susceptible (SIS) model~\cite{keeling08-1}. We note how more realistic setups accounting, for example, for possible latent states of the threats are possible alternatives~\cite{chernikova2023modeling}. In the SIS model, susceptible individuals may receive compromised messages that appear to come from their colleagues, who have already fallen for the deception. The susceptible individuals themselves may then become infected, depending on their level of gullibility. Compromised users eventually realize the issue and recover after a period of time, which also depends on their gullibility. We assume that cyber-threats do not have access to a user’s entire communication history. Instead, they can only piggyback on messages initiated after the user has been compromised and before the threat is detected~\cite{balthrop2004technological}.

In these settings, we assume that, in a given time window, the corporation has a limited budget to increase the awareness of their employees via specific training courses and faces the following question: \textit{who should be selected for such training?} Defining a strategy to select a fraction of employees to better protect a corporation from computer viruses can be formulated as a vaccination problem with limited budget~\cite{newman10-1}. Indeed, users participating to specific training increase their awareness and reduce susceptibility to cyber-threats. For simplicity, we assume that training results in perfect awareness, even though in reality it may be imperfect. As a result, nodes selected for training are removed from the spreading dynamics~\cite{newman10-1}. We investigate the impact of four different strategies. The first acts as a baseline and selects a random fraction of the nodes for training independently from their features. The second focuses on targeting nodes that, possibly due to the nature of their work (e.g., customer service), tend to get in contact with more individuals over time (i.e., they are more active)~\cite{liu2014controlling}. The implementation of this strategy requires a complete knowledge about users' activity. This might be challenging to obtain in reality due to computational costs required to monitor all communications, as well as privacy and ethical constraints. To overcome this limitation, we consider a third strategy based on local egocentric sampling of the connections of a fraction of nodes that act as probes~\cite{liu2014controlling,cohen03-1}. Finally, we study a fourth strategy that targets users based on their knowledge of cyber-threats estimated via prototypical awareness tests (e.g., simulated phishing campaigns)~\cite{jansson2013phishing}. As noted above, Ref.~\cite{prakash10-1} studied a similar question. However, in their settings users' gullibility is considered constant (both in terms of infection and recovery). Furthermore, the immunization strategies they study resemble our second and third approaches, but rely on different aggregated representations of the network.

We derive closed analytical expressions of the epidemic threshold providing insights about the impact of users' features in all four immunization strategies. We quantify the effectiveness of each strategy via large-scale numerical simulations which i) validate the analytical solutions derived and ii) allow the characterization of the dynamics under consideration. Overall, we find a clear hierarchy among strategies in terms of their effectiveness. The activity-based strategy emerges as the most effective. Indeed, selecting nodes based on their activities might reduce the needed fraction of users to halt the spreading by more than one order of magnitude with respect to other strategies. Despite assuming no knowledge about the system, the egocentric strategy emerges consistently as second best. The strategy based on security awareness tests results only marginally better than the baseline. Interestingly, we find that highly homophilic interactions among gullibility classes increase the range of transmissibility parameters that might result in macroscopic outbreaks, but at the same time reduce the reach of cyber-threats confining them within the most gullible group. This result highlight how cyber-threats might survive in rather isolated parts of the networks even if they are not able to spread in most of the others. Identifying these possible breeding grounds might be crucial for elimination campaigns.\\
The paper is organized as follows. In Section~\ref{model} we describe the general structure of the model. In Section~\ref{Immunization_Strategies} we describe the four strategies, present the analytical results, and the numerical simulations. In Section~\ref{Concl} we present our conclusions.

\section{The Model}
\label{model}

In this section, we summarize the main features of the model that acts as the building block of our study. As mentioned, we build on the model proposed in Ref.~\cite{Brett_2019}

We consider a population of $N$ users which exchange directed messages online. They are divided into $Q$ categories describing their susceptibility to cyber-threats (i.e., gullibility classes). Each node features an activity $a$ describing their propensity to initiate communications per unit time. Activities are extracted from a power-law distribution $F(a) \sim a^{-\alpha}$ with $a \in [\epsilon, 1]$. For simplicity, we assume the same distribution of activity across all gullibility classes, i.e.,  $F(a)=F_x(a)$ $\forall x$. The temporal dynamics regulating the interactions between users is the following. At each time step with probability $a\Delta t$ nodes are active. Active nodes select $m$ others and send them a message. The selection is driven by a parameter $p$, which regulates the homophily in the system: with probability $p$ each active node selects at random another user within the same gullibility class. With probability $(1-p)$, instead, the active user sends a message to a user randomly picked among the other classes. At the end of each time step all connections are deleted and the process restarts. 

We describe the propagation of cyber-threats using an SIS model~\cite{keeling08-1}. Hence, users might be compromised (i.e., infected) and recover returning susceptible. Imagine a user that falls for the ruse at time $t$ and realizes to have been compromised taking actions to regain full control of their computer at time $t'$. The threat will attempt to spread further by covertly sending malicious content to all users legitimately contacted between $t$ and $t'$. Each of the contacted users will be infected, thus falling for the ruse, with a probability $\lambda_{x}$ and recover, becoming again susceptible, at rate $\mu_{x}$. As before, $x$ denotes the gullibility class. We stress the asymmetry in the transmission process: an infection can only occur when an infected user contacts a susceptible, and not the opposite. This asymmetry is the key difference with respect to similar models for biological contagion processes.  

As shown in Ref.~\cite{Brett_2019}, in these settings it is possible to compute an analytical expression for the basic reproduction number (i.e., $R_0$) of a cyber-threat unfolding in the system. $R_0$ is defined as the average number of secondary infections generated by a single compromised account in a otherwise susceptible population~\cite{keeling08-1}. The expression for $R_0$ reads: 

\begin{equation}
R_{0}  =  \frac{p \sum_{x} \beta_{x} + \Xi}{\sum_{x} \mu_{x}}
\label{eq:R0}
\end{equation}

Where $\beta_{x}=m\lambda_{x}\langle a\rangle_{x}$ and $\Xi$ is a function of $\beta_{x}$, $\mu_{x}$, and $c_{x, y}$ (i.e., the mixing probabilities among different gullibility classes) and has a specific algebraic expression for a fixed number of classes $Q$. The quantity $\langle a\rangle_{x}$ denotes the average activity in the gullibility class $x$. We refer the reader to Ref.~\cite{Brett_2019} and to the Appendix for more details about the derivation. 

\section{Control strategies}
\label{Immunization_Strategies}

We imagine a hypothetical scenario of a large institution that, in a given time-window, has the budget to provide cyber-security training to a fraction of its workforce. We assume, for simplicity, that the training provides complete protection from future deception-based attacks. The key question is how to select users that will be trained. In these settings, the cyber-security training is equivalent to a sterilizing vaccine~\cite{newman10-1}. Indeed, in our settings, the training reduces the risk of being compromised to zero and users who receive it are removed from the propagation process. As a result, we model the impact of the training as an SIS model where a fraction of the nodes is completely removed from the spreading dynamics.

We consider four strategies to select users for training. In the first one users are selected at random independently from any of their features. The second strategy selects users in decreasing order of activity. The third strategy is based on an egocentric sampling of the network of communication starting from random probes. The fourth targets users based on their knowledge of cyber-threats estimated via prototypical security awareness training (SAT) tests. 

In what follows, we imagine that the training of a fraction $\gamma$ of employees takes place in a given time-window. We then assume that a small fraction of users fall for a deception-based attack and study the impact of each of the four control strategies in hampering the spreading potential of the cyber-threat. In other words, we study how each strategy protects the system from the attack.    

\subsection{Random strategy}

Given the framework discussed in the previous section, in absence of any training, the equation describing the evolution in time of the number of infected nodes with activity $a$ and in class $x$ can be written as:
\begin{eqnarray}
&&d_t I_{a}^{x} = -\mu_{x}I_{a}^{x} + m  \lambda_{x} S_{a}^{x} \times\\ \nonumber
&&
 \left [ p\int da' a' \frac{I_{a'}^{x}}{N^{x}}+(1-p)\sum_{y\neq x} \int da' a' \frac{I_{a'}^{y}}{N-N^{y}} \right ]
\label{eq1}
\end{eqnarray}
In particular, the first term on the right hand side accounts for the recovery process, while the second and third terms in the square brackets account for the possibility of infection due to a compromised message coming, respectively, from inside or outside the gullibility class $x$. In the first strategy, which acts as a baseline, users are randomly selected for training. Hence, we remove a random fraction $\gamma$ of individuals at the beginning of the spreading process: $R_{a}^{x}=\gamma N_a^x$.  During the early stages of the spreading we assume the number of compromised accounts to be small, i.e., $I_a^x\ll N_a^x$. Hence, we can approximate the number of susceptible individuals in each activity class $a$ and gullibility $x$ as $S_{a}^{x} \approx (1- \gamma) N_{a}^{x}$. Eq.~\ref{eq1} then becomes:
\begin{eqnarray}
&&d_{t}I_{a}^{x} = -\mu_{x}I_{a}^{x} + m  \lambda_{x}(1 - \gamma) N_{a}^{x} \times \\ \nonumber
&& \left [ p\int da' a' \frac{I_{a'}^{x}}{N^{x}}+(1-p)\sum_{y\neq x} \int da' a' \frac{I_{a'}^{y}}{N-N^{y}} \right ]
\label{eq2}
\end{eqnarray}
By defining $\lambda_{x}^{rnd}=\lambda_{x}(1 - \gamma)$ and integrating Eq. \ref{eq2} across all activities we obtain: 
\begin{equation}
d_{t}I^{x} = -\mu_{x}I^{x} + m  \lambda_{x}^{rnd}  \left [ p \theta^{x}+(1-p) \sum_{y\neq x} c_{x,y}\theta^{y} \right ]
\label{eq3}
\end{equation}
Where $\theta^{x}=\int da a I_{a}^{x}$, $c_{x,y}=N^{x} / (N-N^{y})$, $I^{x}=\int da I_{a}^{x}$, and $N^{x}=\int da N_{a}^{x}$. By multiplying both sides of Eq.~\ref{eq2} by $a$ and integrating across all activities we obtain:
\begin{equation}
d_{t}\theta^{x} = -\mu_{x}\theta^{x} + m\lambda_{x}^{rnd} \langle a\rangle_{x} \left[ p\theta^{x} + (1-p) \sum_{y\neq x} c_{x,y}\theta^{y} \right]
\label{eq4}
\end{equation}
Equations \ref{eq3} and \ref{eq4} define a system of $2Q$ differential equations. The cyber-threat will be able to spread only if the largest eigenvalue of the Jacobian matrix of the system is greater than zero. As shown in the Appendix, the largest eigenvalue of the system reads:
\begin{equation}
\Lambda_{max}^{rnd} = -\sum_{x}\mu_{x}+p\sum_{x}\beta_{x}^{rnd}+ \Xi^{rnd}
\end{equation}
Where $\beta_{x}^{rnd} = (1-\gamma) \beta_{x}$ and $\Xi^{rnd}$ is a function of $\beta_{x}^{rnd}, \mu_{x}, c_{x,y}$.
Considering the expression of the largest eigenvalue of the system of differential equations, the basic reproduction number in the case of random immunization strategy becomes: 
\begin{equation}
\label{R0_rnd}
R_{0}^{rnd} = \frac{p\sum_{x}\beta_{x}^{rnd} + \Xi^{rnd}}{\sum_{x}\mu_{x}}
\end{equation}
We notice that its expression is equal to the expression of $R_{0}$ without any security training (see Eq.~\ref{eq:R0}), except for the terms $\beta_{x}^{rnd}$ and $\Xi^{rnd}$ that are affected by the definition of $\lambda_{x}^{rnd}$. 
To showcase the full expression of the threshold, we consider the cases with a single and two gullibility classes, i.e., $Q=1$ and $Q=2$.\\

\textbf{Case Q=1}. With a single gullibility class $\Xi^{rnd}=0$ (and $p=1$), hence:
\begin{equation}
R_{0}^{rnd} = \frac{\beta^{rnd}}{\mu} = \frac{(1-\gamma) \beta}{\mu} = (1-\gamma) R_{0} 
\end{equation}
In this case, $R_{0}$ is simply rescaled of a factor $(1-\gamma)$ (i.e., fraction of nodes not removed). This is the classic result of random immunization/removal of nodes. Indeed, the impact of the strategy on the threshold scales linearly with the fraction of nodes removed~\cite{newman10-1}.\\

\textbf{Case Q=2}. With two gullibility classes $(\Xi^{rnd})^{2}$ takes the following expression:
\begin{equation}
\begin{split}
(\Xi^{rnd})^2 = &(\mu_{1} - \mu_{2})^{2} + (1-\gamma)^{2} \times\\
&\left[ \right. p^{2}(\beta_{1}-\beta_{2})^{2} + \frac{2p}{1-\gamma} (\mu_{2} - \mu_{1})(\beta_{1}-\beta_{2}) + \\
&4\beta_{1}\beta_{2}(1-2p)\left. \right]
\end{split}
\end{equation}
If the two gullibility classes feature the same recovery rate (i.e., $\mu_{1} = \mu_{2}$), the expression simplifies to:
\begin{equation}
\begin{split}
(\Xi^{rnd})^2 &= (1-\gamma)^{2}[p^{2}(\beta_{1}-\beta_{2})^{2} + 4\beta_{1}\beta_{2}(1-2p)] \\
&= (1-\gamma)^{2}\Xi^2
\end{split}
\end{equation}
Hence, when the two classes are characterized by the same recovery rate the basic reproduction number $R_0^{rnd}$ is equal to $R_0$ rescaled by factor $(1-\gamma)$. In other words, in case the two gullibility groups differ just by the probability of infection, the impact of a random removal strategy scales linearly with $\gamma$. Interestingly, in case $\mu_{1} \neq \mu_{2}$ this simple relation does not hold anymore. Indeed, in this case there is an interplay between time-scales regulating the infection period of each class. As shown in Ref.~\cite{Brett_2019} in the absence of any intervention strategy, this interplay can make the system more fragile than it would be if each class were considered separately.

\subsection{Activity-based strategy}

This second strategy targets nodes that, possibly due the nature of their job or personal attitude, are more active. Hence, we remove all nodes of class $x$ that feature an activity higher than a given threshold $a_{c}(x)$. In practice, this means that all integrals across activities go from $\epsilon$ to $a_{c}(x)$ (and not $1$). In this case the early stage linearization takes the form $S_{a}^{x} \sim (1 - \gamma_{x}) N_{a}^{x}$ where $\gamma_{x}=\int_{a_{c}(x)}^{1}N_{a}^{x} / N^{x} da$ is the fraction of nodes removed in class $x$. The system of $2Q$ differential equations defined by Eq.~\ref{eq3} and Eq.~\ref{eq4} can be rewritten as:
\begin{equation}
\begin{split}
& d_{t}I^{x} = -\mu_{x}I^{x} + m  \lambda_{x}^{act}  \left [ p \theta^{x}+(1-p) \sum_{y\neq x} c_{x,y}\theta^{y} \right ] \\
& d_{t}\theta^{x} = -\mu_{x}\theta^{x} + m\lambda_{x}^{act} \langle a\rangle_{x}^{c}\left[ p\theta^{x} + (1-p) \sum_{y\neq x} c_{x,y}\theta^{y} \right]
\end{split}
\end{equation}
Where we define $\lambda_{x}^{act}=(1 - \gamma_{x})\lambda_{x}$ and $\langle a\rangle_{x}^{c}=\int_{\epsilon}^{a_{c}(x)} a F_{x}(a)da$.
Following the same steps outline above, we derive the basic reproduction number in the case of activity-targeted immunization strategy as: 
\begin{equation}
R_{0}^{act} = \frac{p\sum_{x}\beta_{x}^{act} + \Xi^{act}}{\sum_{x}\mu_{x}}
\end{equation}
Where $\beta_{x}^{act} = m\lambda_{x}^{act}\langle a\rangle_{x}^{c}$ and $\Xi^{act}$ is a function of all $\beta_{x}^{act}$, $\mu_{x}$, and $c_{x, y}$.\\

\textbf{Case Q=1}. With a single gullibility class $\Xi^{act}=0$ (and $p=1$), hence:
\begin{equation}
R_{0}^{act} = \frac{\beta^{act}}{\mu} = \frac{m\lambda_{x}^{act}\langle a\rangle_{x}^{c}}{\mu} = \frac{m(1 - \gamma)\lambda \langle a\rangle_{x}^{c}}{\mu} 
\end{equation}
In this case, the threshold is not a simple rescaling of the $R_0$ obtained without any interventions. Indeed, the expression is also modified by the contribution of activity classes which are able to get infected. In doing so, the modulation of $R_0$ induced by targeting the most active nodes in each activity class is, generally speaking, not linear with the fraction of nodes removed.\\

\textbf{Case Q=2}. The expression of $\Xi^{act}$ is analogous of $\Xi^{rnd}$ where however the $\beta_x$ are substituted with $\beta_{x}^{act}$. Hence, also in this case, the impact of the selection strategy is regulated by the transmission rates of each class, i.e., $\beta_x^{act}$. As noted above, these are affected by the expression of $\gamma_x$ and the activity distributions of the nodes possibly affected by the threat, i.e., $\langle a\rangle_{x}^{c}$. Hence, the impact of each node removed is again non-linear.  

\subsection{Egocentric sampling strategy}

The activity-based strategy requires a complete knowledge of nodes' activities. Due to practical and privacy issues this information is typically unavailable in real-world scenarios. However, a proxy of nodes' activity can be obtained by sampling the egocentric network of a fraction of nodes~\cite{liu2014controlling,cohen03-1}. Egocentric networks capture the connection that each ego (i.e., a given node in the system) has with their alters (i.e., the first neighbors of each ego). We can sample these egocentric networks by randomly selecting a group of nodes that act as probes. We then observe their connections (i.e., egocentric network) during a time-window of length $T$, neglecting the direction of links. In other words, we sample the interactions of each probe taking place within an observation window. Then, for each of the probes we pick one alter at random in their egocentric network and select it for security training. The idea behind this selection strategy is that highly active nodes are more likely to be in the egocentric network of different probes. This local sampling strategy improves the likelihood of selecting high-activity nodes without assuming any global knowledge about the system.  In general, the number of probes in different classes can vary depending on their size. Given a total number of probes $N_w$, assuming a random distribution, the expected number of these in each gullibility class can be written as $N_{w}^x = N_w \frac{N^x}{N}$. We note how in these settings the fraction of probes in each gullibility class (i.e., $w_x=N_w^x/N_x$) is equal to the total fraction and equal across each class (i.e., $w_x=w$ $\forall x$). However, the number of probes in each class could be different. 

Let us define $P_{a}^{x}$ as the probability that, from the egocentric network of a given probe, we select a node of activity $a$ in the gullibility class $x$. These are the nodes that will undertake the security training and thus will be immune from cyber-attacks. After one observation time step we can write:
\begin{eqnarray}
P_{a}^{x} &=& a p \int da' N_{a'}^x w_x \frac{m}{N_x} + \nonumber\\
&+&a (1 - p) \sum_{y\neq x}\int da' N_{a'}^{y}w_y \frac{m}{N-N_x}+ \nonumber\\
&+&\int da' a'p N_{a'}^{x}w_x\frac{m}{N_x}\frac{1}{m} \nonumber\\ 
&+&\sum_{y\neq x} \int da' a'(1 - p)N_{a'}^{y}w_y \frac{m}{N-N_y} \frac{1}{m} \nonumber\\
&=& apw_xm+a(1-p)m\frac{N_w-N_w^x}{N-N_x}+pw_x\langle a \rangle_x \nonumber\\
&+&(1-p)\sum_{y\neq x}\frac{N^y}{N-N_y} w_y \langle a \rangle_y
\label{eq:egocentricprob}
\end{eqnarray}
In particular, the first term of Eq.~\ref{eq:egocentricprob} represents the probability that nodes with activity $a$ and in gullibility class $x$ are selected for training (i.e., are removed from the cyber-threat dynamics) because they are active and connect with probes in the same gullibility class; the second term is analogous but considers connections with probes in other gullibility classes; the third and the fourth term, instead, represent the probability that nodes are removed after being reached and selected from probes, respectively, inside (third term) and outside (forth term) their gullibility class. By assuming this selection dynamics independent across time steps, the probability for a node with activity $a$ and in class $x$ to be removed after $T$ time-steps can be written as $P_{a}^{x}(T) = 1 - (1-P_{a}^{x})^{T}$. Hence, the number of nodes removed after $T$ periods with activity $a$ and in class $x$ is $R_{a}^{x}(T)= N_{a}^{x}(1 - (1-P_{a}^{x})^{T})$.
We note how this formulation is a clear approximation. Indeed it does not consider the depletion of nodes in each class due to the immunization process. As such, this expression holds in the regime of small $T$ and when the probability that a probe is selected more than once is small. Furthermore, we note how due to the possible selection of the same targets from different probes, in general, the cardinality of the set of nodes selected by this strategy for security training, $\gamma$, might be smaller than the fraction of probes $\gamma \le w$. As before, at early stages of the spreading, we can write $S_{a}^{x} \sim N_{a}^{x}-R_{a}^{x}(T)$ and repeat similar calculations to those explained above to obtain:
\begin{equation}
\begin{split}
d_{t}I^{x} = -\mu_{x}I^{x} + &m  \lambda_{x} \Psi_{0,x}^{T} \times\\
&\left [ p\theta^{x}+(1-p)\sum_{y\neq x} c_{x,y} \theta^{y} \right ] 
 \end{split}
 \end{equation}
 
 \begin{equation}
\begin{split}
d_{t}\theta^{x} = -\mu_{x}\theta^{x} + &m  \lambda_{x}\Psi_{1,x}^{T}\times\\
 &\left [ p\theta^{x}+(1-p)\sum_{y\neq x} c_{x,y} \theta^{y} \right ] 
 \end{split}
 \label{eq:thetaego}
\end{equation}
Where we define $\Psi_{n,x}^{T} = \int da a^{n} F_{x}(a)(1-P_{a}^{x})^{T}$. In this case, the basic reproduction number can be written as: 
\begin{equation}
R_{0}^{ego} = \frac{p\sum_{x}\beta_{x}^{ego} + \Xi^{ego}}{\sum_{x}\mu_{x}} \quad \text{with} \quad \beta_{x}^{ego} = m \lambda_{x} \Psi_{1,x}^{T}
\end{equation}\\

\textbf{Case Q=1}. In case of a single gullibility class we have $R_{0}^{ego}=\frac{\beta^{ego}}{\mu}=\frac{m\lambda}{\mu}\int da a F(a)(1-P_{a})^{T}$. We note how the egocentric sampling strategy affects the threshold by decreasing, non-linearly as function $T$, the average activity of susceptible nodes. \\

\textbf{Case Q=2}. In case of two gullibility classes the expression of $\Xi^{ego}$ is analogous of $\Xi^{rnd}$ where $\beta^{rnd}_x$ are substituted with $\beta_{x}^{ego}$. Also in this case, the impact of strategy on the dynamics is hidden in the $\Psi_{1,x}^{T}$ expressions which lead to non-linear effects.

\subsection{Security Awareness Training Strategy}
In this last strategy, we imagine that the corporation runs a security awareness training (SAT) test in which all the employees (i.e., nodes) receive a fake compromised email and/or message (e.g., a phishing email). These tests are customarily used for cyber-security training and awareness purposes~\cite{al2017security}. The strategy consists in estimating the gullibility of employees based on the outcomes of the test. In particular, we implement it as follows. With probability $g$, a user sees the SAT email and opens it. In general, we set $g<1$ thus not all employees engage with the SAT. After seeing the email, a node in gullibility class $x$ clicks on the compromised link, thus falling for the ruse, with probability $\lambda_x$. We assume that the fraction $\gamma$ of the users that are selected to receive security training is selected from the pool of employees that did not recognize the potential threat and clicked on the compromised link. In doing so, we aim to select users more in need of security training.

In these settings, the average number of employees with activity $a$ and in gullibility class $x$ that would fall for the ruse can be estimated as $gN_a^x \lambda_x$. The fraction $\gamma'$ of these needed such that the overall fraction of employees ultimately selected for training is $\gamma$ can be obtained solving the following equation $\gamma=N^{-1} \sum_x \int da gN_a^x \lambda_x \gamma'$. This leads to: 
\begin{equation}
\gamma' = \frac{\gamma}{g \langle \lambda \rangle},
\end{equation}
where $\langle \lambda \rangle = \sum_x \lambda_x N_x / N$ is the average transmissibility in the system. Hence, the number of employees with activity $a$ and in gullibility class $x$ selected for training can be written as: 

\begin{equation}
R_a^x = gN_a^x \lambda_x \gamma'=N_a^x \frac{\lambda_x}{\langle \lambda \rangle} \gamma.
\end{equation}

Interestingly, if a class $x$ features transmissibility equal to the network's average, the fraction of removed nodes in that class is simply the one of the random case (i.e., $R_a^x \sim N_a^x \gamma$). If a class $x$ has transmissibility higher (lower) than the average (i.e., is more or less gullible than the average), it will have a higher (lower) fraction of removed nodes with respect to the random case. 

At early stages of the spreading, the equation for $I_a^x$ becomes:

\begin{eqnarray}
&&d_t I_{a}^{x} = -\mu_{x}I_{a}^{x} + m  \lambda_{x} N_{a}^{x} \left (1 - \frac{\lambda_x}{\langle \lambda \rangle} \gamma \right ) \times\\ \nonumber
&&
 \left [ p\int da' a' \frac{I_{a'}^{x}}{N^{x}}+(1-p)\sum_{y\neq x} \int da' a' \frac{I_{a'}^{y}}{N-N^{y}} \right ]
\end{eqnarray}

By defining $\lambda_x^{sat} = \lambda_{x}\left (1 - \frac{\lambda_x}{\langle \lambda \rangle} \gamma \right )$, we obtain an equation analogous to the random case. Hence, we can directly write the expression for $R_0$ as: 
\begin{equation}
R_{0}^{sat} = \frac{p\sum_{x}\beta_{x}^{sat} + \Xi^{sat}}{\sum_{x}\mu_{x}}, 
\end{equation}
where $\beta_{x}^{sat} = \beta_x \left (1 - \frac{\lambda_x}{\langle \lambda \rangle} \gamma \right )$ and $\Xi^{sat}$ is again a function of $\beta_{x}^{sat}$, $\mu_x$, $c_{x,y}$.\\

\textbf{Case Q=1}. In case of a single gullibility class the SAT strategy is equivalent to the random strategy. Indeed, in this case nodes are selected proportionally to their gullibility and not other features. Thus in case of a single group of nodes, each one is selected uniformly at random. This aspect of the SAT strategy hints to its difference with respect to the previous two strategies which, even in the case of one gullibility class, did not lead to the same expression of the baseline strategy. \\

\textbf{Case Q=2}. In case of two gullibility classes the expression of $R_0^{sat}$ is analogous to the random case. However, as for the other cases, the expression of the $\beta$ terms is different. The effect of the selection strategy is function of $\gamma$ and modulated by the gullibility of each class with respect to the system's average.\\
Interestingly, as shown in the Appendix, for any number of gullibility classes and a given fraction of removed nodes $\gamma$, the effective average transmissibility in this strategy cannot be larger than in the random case, namely $\langle \lambda^{sat} \rangle \le \langle \lambda^{rnd} \rangle$. This implies that the effective spreading potential in case the subset of nodes is selected via a SAT strategy can be only smaller or equal with respect to a random selection. \\

\subsection{Numerical simulations}

\begin{figure}[t!]
  \centering
  \includegraphics[width=\linewidth]{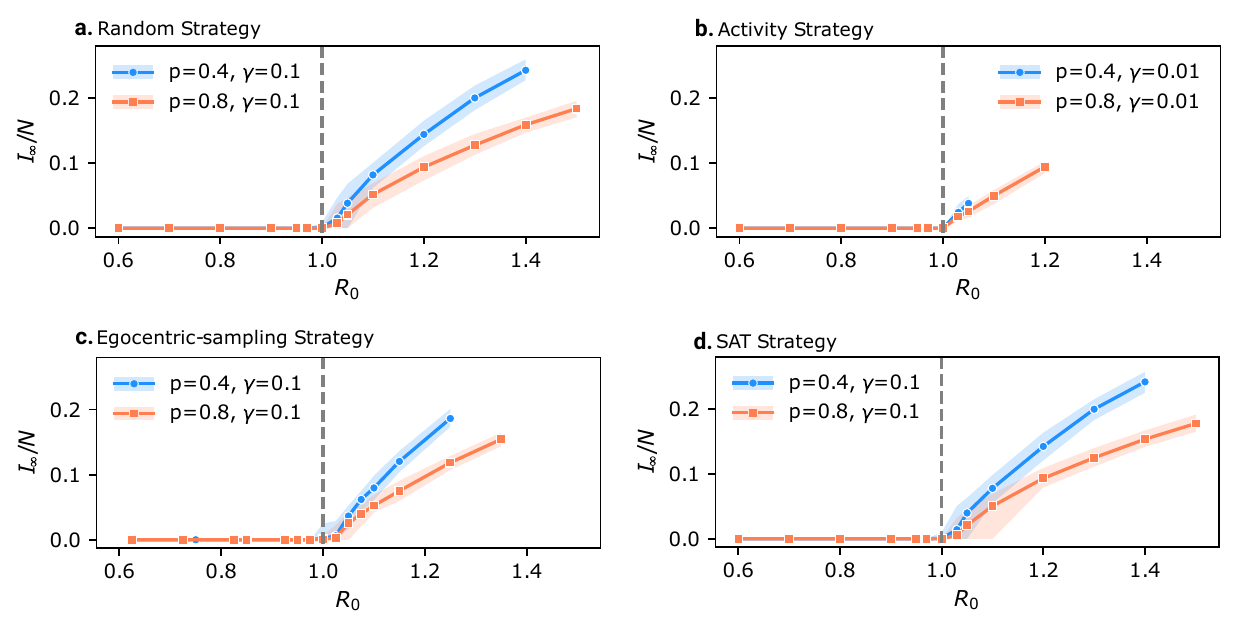}
  \caption{\textbf{Numerical validation of the threshold under different strategies}. Each panel shows the stationary fraction of infected individuals $I_\infty / N$ as a function of $R_0$, for the four immunization strategies: (a) Random, (b) Activity-based, (c) Egocentric sampling, and (d) SAT. Results are shown for two parameter settings: $(p=0.4, \gamma=10^{-1}, \lambda_2 = 0.5)$ in blue and $(p=0.8, \gamma=10^{-1}, \lambda_2 = 0.3)$ in orange for panels (a), (c), and (d); and $(p=0.4, \gamma=10^{-2}, \lambda_2 = 0.5)$ in blue and $(p=0.8, \gamma=10^{-2}, \lambda_2 = 0.8)$ in orange for panel (b). The vertical dashed line indicates the critical value of the threshold computed analytically (i.e., $R_0 = 1$). Solid lines with markers represent mean values while shaded areas indicate $95\%$ confidence intervals computed in $100$ stochastic simulations. Other parameters common to all simulations: $\alpha=2.1$, $\epsilon=10^{-3}$, $m=4$, initial infected percentage $0.5\%$, $\mu_1=\mu_2=10^{-2}$. In the egocentric sampling strategy case we set $T=10$.}
  \label{fig:fig1}
\end{figure}

In Fig.~\ref{fig:fig1} we show, for each strategy, the simulated fraction of infected nodes at the equilibrium as a function of $R_0$ for two different values of $p$ and two gullibility classes ($Q=2$). In all simulations, we assume the process reaches the stationary state when the ratio between the mean and the standard deviation of the prevalence (i.e., number of currently infected nodes), computed over the last $10^3$ simulation steps, falls below a threshold of $0.02$. In all scenarios, except those adopting an activity-based strategy, we set $\gamma=10^{-1}$ (i.e., $10\%$ of employees are enrolled in the security training). The strategy that selects nodes in decreasing order of activity is so effective that, to validate the analytical formulation, we need to consider smaller values of $\gamma$ (e.g., $\gamma=10^{-2}$). Indeed, all physical combinations of parameters lead to sub-critical states for $\gamma=10^{-1}$. In all simulations, we fixed the infection probability of the second class (i.e., $\lambda_2$) and let $\lambda_1$ vary exploring corresponding $R_0$ values in the range $0.6$ to $1.5$. We exclude non-physical combinations that result in values $\lambda_1 > 1$. Furthermore, we consider a simple case in which the two recovery rates are equal, i.e., $\mu_1=\mu_2$. As a way to show the validity of the analytical derivation across a wider range of parameters, we set two different values of $\lambda_2$ for the two values of $p$. We use $\lambda_2=0.3$ for $p=0.8$ and $\lambda_2=0.5$ for $p=0.4$. We note how we adopt $R_0$ as order parameter rather than $\lambda_1$ to fairly compare different parameters combinations. The analytical estimation of the thresholds for all strategies is confirmed by the numerical simulations. Indeed, the analytical thresholds clearly split the phase spaces in two. Below the critical value (i.e., $R_0=1$) the cyber-threat is not able to spread into the system. Then, to the right of the critical values, we see a clear transition in the dynamics. Indeed, the fraction of infected nodes reaches an endemic state. The effectiveness of each strategy can be evaluated by looking at the outbreak size for a given $R_0$. Differences become clear as we move away from the threshold (i.e., $R_0=1$). The activity-based strategy emerges are clearly the most effective. Indeed, the values $\lambda_1$ above threshold are just extreme values very close to $1$. This explains why we have fewer points in that panel (see Fig.~\ref{fig:fig1}-b). The effectiveness of this strategy is even more striking recalling than in this case we removed only $1\%$ of nodes, rather than $10\%$ as for the other strategies. The baseline and the SAT strategy appear similar and clearly less effective than the activity-based. As mentioned above, the similarity between the two is to be expected by construction. However, the SAT strategy performs marginally better, especially for larger values of $p$. The egocentric strategy appears to be more effective than both the baseline and the SAT strategy. However, its performance is still far from the most performant. Across the board, we observe how smaller values of $p$ (i.e., low homophily) result in larger outbreaks especially for large values of $R_0$. Hence, well above the threshold, increased mixing across gullibility classes might be detrimental to the whole system in case of successful attacks.

\begin{figure}[t!]
  \centering
  \includegraphics[width=\linewidth]{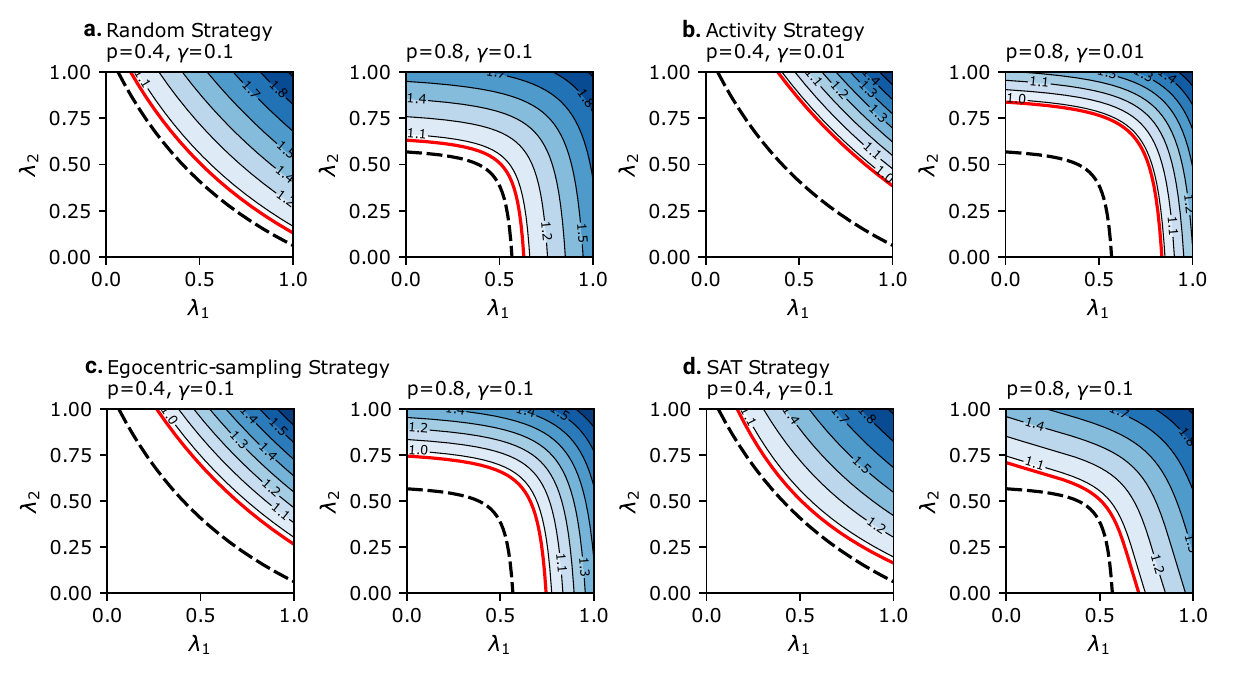}
  \caption{\textbf{Phase space of $R_0$ as a function of $\lambda_1$ and $\lambda_2$, under different strategies}. Each panel corresponds to a specific strategy: (a) Random, (b) Activity-based, (c) Egocentric sampling, and (d) SAT, with two sub-panels per strategy representing $(p=0.4, \gamma=10^{-1})$ and $(p=0.8, \gamma=10^{-1})$—except for panel (b), which uses $\gamma=10^{-2}$. The colored contours show analytically computed $R_0$ values for each strategy. The red solid contour line marks the critical threshold $R_0 = 1$ under the given intervention strategy, while the black dashed line shows the $R_0 = 1$ threshold in the absence of intervention. Other parameters common to all panels: $\mu_1=\mu_2=10^{-2}$, $m=4$, $\alpha=2.1$, $\epsilon=10^{-3}$.}
  \label{fig:fig2}
\end{figure}

In Fig.~\ref{fig:fig2} we show contour plots of the theoretical value of $R_0$, estimated from the analytical derivations described above, as a function of the gullibility of the two classes, i.e., $R_0(\lambda_1, \lambda_2)$. As for the previous plots, we assume $\mu_1=\mu_2$. The black dashed lines show the thresholds (i.e., $R_0=1$) in case of $\gamma=0$ (no security training). The red solid lines, instead, show the thresholds in case of $\gamma=10^{-1}$ for all strategies but the activity-based one. As before, we set $\gamma=10^{-2}$ for this strategy. In each panel, the gap between the two lines quantifies the impact of the training strategy. Indeed, points below each line are subcritical, thus the threat would not be able to spread in those regions of the phase space. The two lines are rather close in the case of the random baseline strategy highlighting the marginal efficacy of this strategy (see Fig.~\ref{fig:fig2}-a). The effectiveness of the activity-based strategy clearly emerges in Fig.~\ref{fig:fig2}-b. Indeed, the gap between the two lines is the largest among all strategies, confirming how selecting nodes based on their activities leads to the best outcomes. We stress one more time how the effectiveness of this strategy is particularly striking when considering that it is the only one for which we removed only $1\%$ of nodes. The egocentric strategy is confirmed more effective than both random and SAT strategies with a gap between the two lines closer to the activity-based strategy, though in this case we have $\gamma=10^{-1}$. The SAT strategy is confirmed similar to the random baseline (see Fig.~\ref{fig:fig2}-d), though more effective, especially for larger values of homophily and when the two classes exhibit greater differences in gullibility. Across all strategies, the difference of the phase spaces as function of $p$ shows how large values of homophily allow macroscopic, yet localized, outbreaks even if one of the two classes is perfectly immune to the threat (e.g., $\lambda_1=0$). Indeed, in these scenarios the threat is able to spread, and survive, in one community of the network. On the other hand, smaller values of homophily lead to a larger mix between the two classes and dynamics is driven by the interplay between the gullibility of the two classes. By observing the critical value of $\lambda_2$ above which the threat would be able to spread even if the other gullibility class is perfectly protected (i.e., $\lambda_1=0$) offers another approach to compare strategies. Indeed, higher values the $\lambda_2$ highlight better performance in stopping the spreading. This value is the largest in the case of the activity-based strategy ($\lambda^c_2\simeq 0.84$). The egocentric strategy follows with a critical value of $\lambda^c_2 \simeq 0.75$. The SAT and random strategy then shows values of $\lambda^c_2 \simeq 0.71$ and $\lambda^c_2 \simeq 0.63$ respectively.

\begin{figure}[t!]
  \centering
  \includegraphics[width=\linewidth]{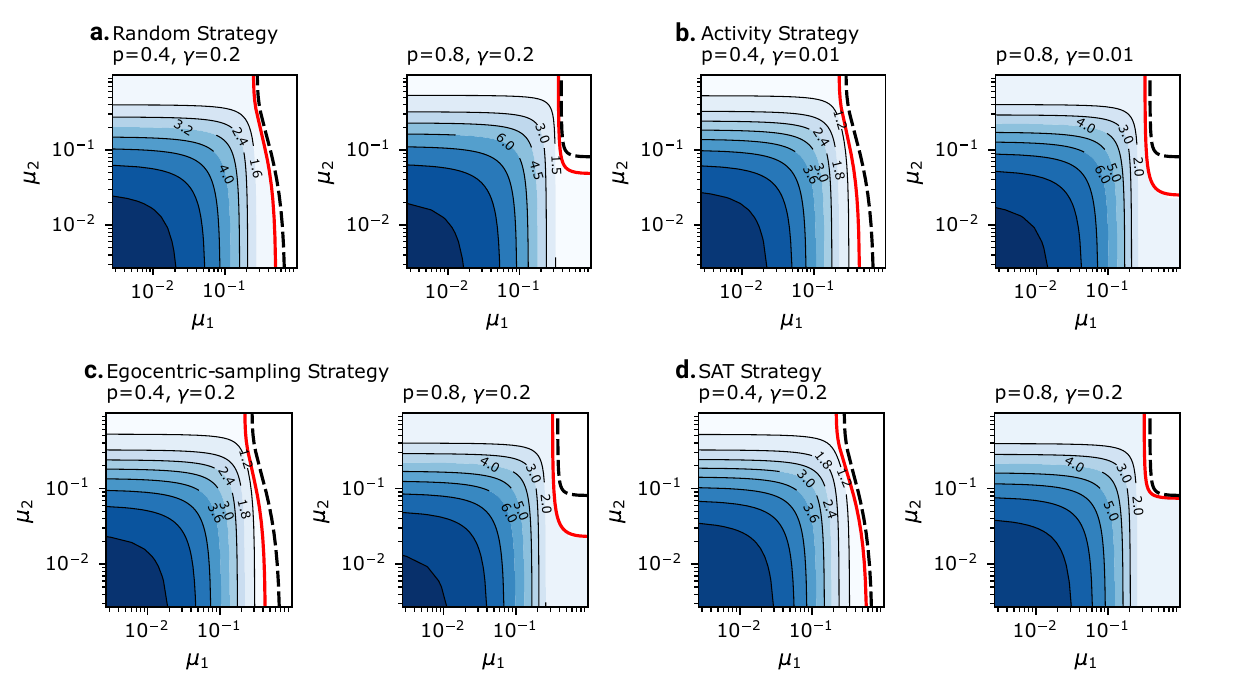}
  \caption{\textbf{Phase diagrams of $R_0$ as a function of $\mu_1$ and $\mu_2$, under different immunization strategies}. Each panel corresponds to a specific strategy: (a) Random, (b) Activity-based, (c) Egocentric-sampling, and (d) SAT, with two sub-panels per strategy representing $(p=0.4, \gamma=0.2)$ and $(p=0.8, \gamma=0.2)$—except for panel (b), which uses $\gamma=10^{-2}$. The colored contours show analytically computed $R_0$ values for each strategy. The red solid contour line marks the critical threshold $R_0=1$ under the given intervention strategy, while the black dashed line shows the $R_0=1$ threshold in the absence of intervention. Other parameters common to all panels: $\lambda_1=10^{-1}$, $\lambda_2=0.8$, $m=4$, $\alpha=2.1$, $\epsilon=10^{-3}$.}
  \label{fig:fig3}
\end{figure}

In Fig.~\ref{fig:fig3} we show the phase space of $R_0$ as function of $\mu_1$ and $\mu_2$ while fixing the values of $\lambda_1$ and $\lambda_2$. We fix $\gamma$ in each case setting $\gamma=0.2$ for all strategies but for the activity-based strategy where we use instead $\gamma=10^{-2}$. Across the board we observe that smaller values of recovery rates result in larger $R_0$. The trend is to be expected as, the larger recovery time (i.e., $\mu_x^{-1}$), the higher the number of opportunities for each infected node to spread the threat further. Also in this plot we observe how large values of homophily allow macroscopic outbreaks even if one gullibility class manages to recover immediately after infection (i.e., $\mu_x=1$), thus limiting the spread of the threat. Furthermore, the plots confirm the hierarchy of efficacy of the four strategies highlighted above. 

\begin{figure}[t!]
  \centering
  \includegraphics[width=\linewidth]{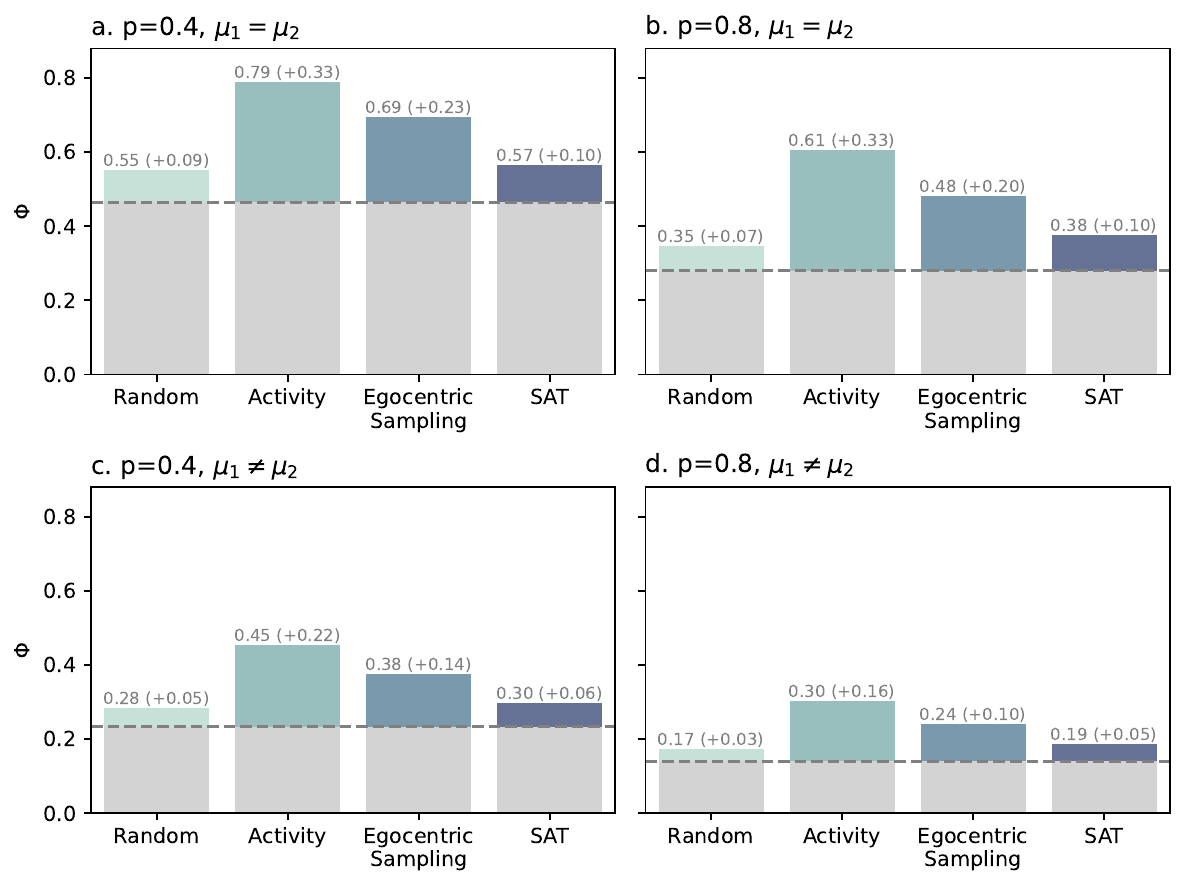}
  \caption{\textbf{subcritical $(\lambda_1, \lambda_2)$ phase space fraction ($\Phi$) under different strategies}. Each panel reports results for a specific combination of parameters: (a) $p = 0.4$, $\mu_1 = \mu_2 = 10^{-2}$; (b) $p = 0.8$, $\mu_1 = \mu_2 = 10^{-2}$; (c) $p = 0.4$, $\mu_1 = 10^{-2}$, $\mu_2 = 5 \times 10^{-2}$; (d) $p = 0.8$, $\mu_1 = 10^{-2}$, $\mu_2 = 5 \times 10^{-2}$. The horizontal dashed lines and gray bars represent the value of $\Phi$ in the absence of intervention. Numerical labels above the bars indicate the total controlled fraction and, in parentheses, the gain with respect to the no-intervention baseline. Other parameters common to all panels: $m=4$, $\alpha=2.1$, $\epsilon=10^{-3}$.}
  \label{fig:fig4}
\end{figure}

To further compare the various strategies, we compute the fraction $\Phi$ of the $(\lambda_1, \lambda_2)$ phase space that results in subcritical dynamics. In other words, $\Phi$ is defined as the portion of the $(\lambda_1, \lambda_2)$ phase space for which the corresponding $R_0$ is subcritical (i.e., $R_0 < 1$). For a given range of parameters, the larger $\Phi$, the smaller region of the transmissibility parameters that would allow for a macroscopic outbreak. We show the results for different scenarios in Fig.~\ref{fig:fig4}. In all plots, the dashed horizontal lines and grey bars describe the value of $\Phi$ in absence of any control strategy (i.e., $\gamma=0$). Furthermore, the numerical labels above each bar indicate the value of $\Phi$ for each strategy and, in parentheses, the absolute gain with respect to the no-intervention scenario. The panels in the first row are obtained considering the same value of the recovery rates in two gullibility classes, but two different values of the homophily parameter. The second row instead consider scenarios in which the recovery parameters are different. A few observations are in order. First, across the board, the hierarchy of effectiveness of the four strategies confirms previous findings. The activity-based strategy results in the largest increase of $\Phi$. Second, larger values of homophily result in smaller subcritical regions of the phase space. Indeed, as observed above, for large values of $p$ the threat might be able to spread even if one gullibility class is perfectly protected (e.g., $\lambda_1=0$). These configurations are not compatible with macroscopic outbreaks in case of higher mixing levels between gullibility classes (i.e., smaller homophily). We note how these results are not in contrast with the observations we made above where we noted how, for a given value of $R_0$, higher levels of homophily corresponded to larger outbreaks. Indeed, $\Phi$ quantifies the inactive (i.e., subcritical) region of the phase space. It is agnostic to the prevalence/reach of the cyber-threat in the system. It is a measure of the combination of transmissibility parameters that result in subcritical dynamics. It does not provide any information about what happens in supercritical regimes. These two results together suggest how higher values of homophily facilitate the spreading of cyber-threats, but limit the their reach in rather isolated groups. Hence, cyber-threats might survive in patches of the network constituted by isolated and highly gullible groups. Third, increasing the recovery rate of even just one class, leads to a sensible reduction of the subcritical phase space. Indeed, the values of $\Phi$ decrease across the board in plots Fig.~\ref{fig:fig4}-c and Fig.~\ref{fig:fig4}-d. We note however how the relative effectiveness of each strategy is preserved also in this case.  

\section{Conclusions}
\label{Concl}
We studied the effectiveness of different strategies aimed at containing the spread of deception-based cyber-threats in online social networks. To this end, we modeled the temporal interactions among users using the framework of activity-driven networks. We allowed for the presence of multiple gullibility (i.e., susceptibility) classes describing heterogeneous risk profiles of users. Furthermore, we assumed that the membership to a gullibility class affects the interaction dynamics via a tunable homophily parameter. Finally, we simulated the spreading of cyber-threats using prototypical SIS epidemic models. In these settings, we quantified the efficacy of four strategies aimed at selecting a fraction of nodes to be protected from such threats. The first strategy acts as a baseline and selects individuals at random. The second assumes complete knowledge of the activity (i.e., propensity of initiating online interactions) of each individual and targets first the most active nodes. The third is based on an egocentric sampling strategy aimed at reaching highly active nodes without assuming any knowledge about their activity. The fourth is based on estimating the gullibility of each node via security awareness tests which are routinely employed in many institutions to probe the cyber-security awareness of the workforce~\cite{al2017security}. We analytically derived the epidemic threshold under each intervention strategy. In doing so, we quantified their effectiveness to control the spreading process. Large-scale numerical simulations validated the analytical expressions across all strategies. The results obtained clearly show the high effectiveness of activity-based strategies which are able to outperform the others even when protecting a smaller fraction of individuals. The egocentric sampling strategy emerges as second best, confirming the value of local sampling strategies aimed at reaching the most active nodes without global knowledge of the system. The fourth strategy based on security awareness tests proved only marginally better than the baseline.\\
Our findings highlight that highly homophilic interactions within gullibility classes expand the transmissibility phase space, thereby fostering conditions for macroscopic outbreaks. Indeed, in these conditions the cyber-threat may still spread within the most gullible group, even when it cannot propagate through others. At the same time, we find that larger values of homophily ultimately reduce the outbreak size with respect to more mixed scenarios. The modulation effects induced by the mixing levels between gullibility classes highlight the importance of considering heterogeneous susceptibility groups. Indeed, neglecting them in favor of a homogeneous representation of gullibility might lead to misrepresentation of the spreading potential of cyber-threats and of the efficacy of strategies aimed at hampering them. Furthermore, these results suggest how cyber-threats might survive and propagate in rather isolated groups of gullible individuals and highlight the importance of identifying and increasing the awareness of these communities.\\ 
The work presented comes with several limitations. First, we neglected more realistic mechanisms driving the interaction between users. Indeed, while we accounted for homophily, we did not consider popularity and social reinforcement mechanisms among others~\cite{sun2023effects,holme2015modern}. Second, for simplicity we assumed indefinite and perfect protection granted by cyber-security training. Third, we considered the recovery process as a spontaneous transition function of the gullibility of each node. Hence, we did not account for the possibility that a compromised account might be informed by others, in response to their anomalous behavior. Fourth, we used a simple SIS compartmentalization setup to model cyber-threats. Fifth, we did not account for possible correlations between activity and gullibility. Finally, we modeled the probability of falling for a ruse and getting infected to be a function only of the gullibility class of each node. Hence, we neglected possible modulations induced by past experiences (i.e., past infection events), recency, and frequency biases~\cite{cranford2021towards}. We leave accounting for these limitations to future work.\\
Overall, our results highlight the striking effectiveness of targeted strategies based on node activity. At the same time, they confirm the effectiveness of local sampling strategies that, although not as performant as targeted approaches, do not require access to global information about systems' connections. The research contributes to the limited literature devoted to controlling the spread of cyber-threats accounting for both temporal dynamics and heterogeneous susceptibility of users.\\

This material is based upon work supported by, or in part by, the U.S. Army Research Laboratory and the U.S. Army Research Office under contract/grant number W911NF2410169. The authors thank G. Loukas for insightful conversations.

\section*{Appendix}

\subsection*{Spreading threshold derivation for $\gamma=0$}
\label{sec:sis_model}

The interactions between nodes follow the model proposed in Ref.~\cite{Brett_2019}. A population of $N$ users is divided into $Q$ categories describing their susceptibility to cyber-threats (i.e. gullibility classes). Nodes feature an activity $a$ describing their propensity to initiate communications. Activities are extracted from a power-law distribution $F(a) \sim a^{-\alpha}$ with $a \in [\varepsilon, 1]$. In these settings, at each time step $t$ a network is generated as follows:
\begin{enumerate}
    \item Each node is initially disconnected.
    \item  With probability $a\Delta t$ each node becomes active.
    \item Active nodes select $m$ others and create directed links (e.g., send them a message). Furthermore, with probability $p$ the new links are created within the same class at random. With probability $1-p$ links are created with nodes in other gullibility classes, at random.
    \item At time $t+\Delta t$ all links are deleted and the process re-starts.
\end{enumerate} 
Without lack of generality, we can set $\Delta t=1$. \\

We simulate the spreading of deception-based cyber-threats unfolding on top of these temporal networks using a classic Susceptible-Infected-Susceptible (SIS) model~\cite{keeling08-1}. Nodes in class $x$ get infected with probability $\lambda_x$ and spontaneously become again susceptible at rate $\mu_x$. We stress the asymmetry in the transmission process: an infection can only occur when an infected contacts a susceptible, and not the opposite.\\

Assuming that all nodes in the same gullibility and activity class are statistically equivalent, and considering the continuous limit (i.e., $N\rightarrow \infty$), we can write the equation describing the evolution of the number of infected as:

\begin{widetext}
\begin{equation}
d_tI^x_a = -\mu_x I^x_a + \lambda_x m S^x_a \left[ p \int da' a' \frac{I^x_{a'}}{N^x} + (1 - p) \sum_{y \ne x} \int da' a' \frac{I^y_{a'}}{N - N^y} \right]
\label{eq:iax}
\end{equation}
\end{widetext}

The first term of the right hand side captures the recovery process. The second term describes susceptible nodes that receive a compromised message coming from their gullibility class and as result get infected. The third is analogous to the previous term but accounts for compromised messages arriving from other gullibility classes. At early stages, we can assume that the number of compromised nodes is very small, hence we can consider the approximation $S^x_a \sim N^x_a$. The previous equation becomes:

\begin{widetext}
\begin{equation}
d_tI^x_a = -\mu_x I^x_a + \lambda_x m N^x_a \left[ p \int da' a' \frac{I^x_{a'}}{N^x} + (1 - p) \sum_{y \ne x} \int da' a' \frac{I^y_{a'}}{N - N^y} \right]
\label{eq:iax_early}
\end{equation}
\end{widetext}

We observe that $\int da I^x_a = I^x$, $\int da N^x_a = N^x$. Integrating both members of Eq.~\ref{eq:iax_early} over all activities we obtain:

\begin{equation}
d_tI^x = -\mu_x I_x + \lambda_x m \left[ p\theta^x + (1 - p) \sum_{y \ne x} c_{x,y} \theta^y \right]
\end{equation}

Where we define $\theta^x = \int da' a I^x_a$, $c_{x,y} = N^x / (N - N^y)$. We multiply both members of Eq.~\ref{eq:iax_early} by $a$ and integrate over all activities:

\begin{equation}
d_t\theta^x = -\mu_x \theta^x + \lambda_x m \langle a \rangle_x \left[ p\theta^x + (1 - p) \sum_{y \ne x} c_{x,y} \theta^y \right]
\end{equation}

Where we define $\langle a \rangle_x = \int da a N^x_a / N^x$. Finally, we obtain a system of $2Q$ differential equations that describes the evolution of the system:

\begin{align}
d_tI^x &= -\mu_x I^x + \lambda_x m \left[ p\theta^x + (1 - p) \sum_{y \ne x} c_{x,y} \theta^y \right] = g^x \nonumber \\
d_t\theta^x &= -\mu_x \theta^x + \lambda_x m \langle a \rangle_x \left[ p\theta^x + (1 - p) \sum_{y \ne x} c_{x,y} \theta^y \right] = h^x
\end{align}

The threat would be able to spread if the largest eigenvalue of the Jacobian matrix of this system is larger than zero. The Jacobian matrix can be written as:

\[
J = \begin{bmatrix}
\frac{\partial g^1}{\partial I^1} & \cdots & \frac{\partial g^1}{\partial I^Q} & \frac{\partial g^1}{\partial \theta^1} & \cdots & \frac{\partial g^1}{\partial \theta^Q} \\
\vdots & \ddots & \vdots & \vdots & \ddots & \vdots \\
\frac{\partial g^Q}{\partial I^1} & \cdots & \frac{\partial g^Q}{\partial I^Q} & \frac{\partial g^Q}{\partial \theta^1} & \cdots & \frac{\partial g^Q}{\partial \theta^Q} \\
\frac{\partial h^1}{\partial I^1} & \cdots & \frac{\partial h^1}{\partial I^Q} & \frac{\partial h^1}{\partial \theta^1} & \cdots & \frac{\partial h^1}{\partial \theta^Q} \\
\vdots & \ddots & \vdots & \vdots & \ddots & \vdots \\
\frac{\partial h^Q}{\partial I^1} & \cdots & \frac{\partial h^Q}{\partial I^Q} & \frac{\partial h^Q}{\partial \theta^1} & \cdots & \frac{\partial h^Q}{\partial \theta^Q}
\end{bmatrix}
\]

Substituting the partial derivatives, we get a block matrix whose structure depends on $\mu_x$, $\lambda_x$, $c_{x,y}$, $\langle a \rangle_x$, and $p$:

\begin{widetext}
\[
J =
\left[
\begin{array}{cccc|ccc|ccc}
-\mu_1 & 0 & \cdots & 0 & p\lambda_1 m & (1-p)\lambda_1 m c_{1,2} & \cdots & (1-p)\lambda_1 m c_{1,Q} \\
0 & -\mu_2 & \cdots & 0 & (1-p)\lambda_2 m c_{2,1} & p\lambda_2 m & \cdots & (1-p)\lambda_2 m c_{2,Q} \\
\vdots & \vdots & \ddots & \vdots & \vdots & \vdots & \ddots & \vdots \\
0 & 0 & \cdots & -\mu_Q & (1-p)\lambda_2 m c_{Q,1} & (1-p)\lambda_2 m c_{Q,2} & \cdots & p\lambda_Q m \\
\hline
0 & 0 & \cdots & 0 & -\mu_1 + p \beta_1 & (1-p)\beta_1 c_{1,2} & \cdots & (1-p)\beta_1 c_{1,Q} \\
0 & 0 & \cdots & 0 & (1-p)\beta_2 c_{2,1} & -\mu_2 + p \beta_2 & \cdots & (1-p)\beta_2 c_{2,Q} \\
\vdots & \vdots & \ddots & \vdots & \vdots & \vdots & \ddots & \vdots \\
0 & 0 & \cdots & 0 & (1-p)\beta_Q c_{Q,1} & (1-p)\beta_Q c_{Q,2} & \cdots & -\mu_Q + p \beta_Q
\end{array}
\right]
\]
\end{widetext}

Where we define $\beta_x = m \langle a \rangle_x \lambda_x$. Then, the largest eigenvalue $\Lambda_{\max}$ can be written in general form as~\cite{Brett_2019}:

\begin{equation}
\Lambda_{\max} = -\sum_x \mu_x + p \sum_x \beta_x + \Xi
\end{equation}

Where $\Xi$ is an algebraic function of $\beta_x$, $\mu_x$, and $c_{x,y}$ and its analytical expression depends from the number of classes $Q$. Since the cyber-threat is able to spread if $\Lambda_{\max}>0$, we define the basic reproduction number as:

\begin{equation}
R_0 = \frac{p \sum_x \beta_x + \Xi}{\sum_x \mu_x}
\end{equation}

If $R_0 > 1$ the threat would be able to spread affecting a macroscopic fraction of the population.

\subsection*{Random strategy}

Here, we provide the details about the threshold derivation for the random strategy. We recall that in this case the fraction of removed nodes, $\gamma$ is picked at random, independently of any of their features. Hence, in the early stages of the spreading we can write:

\begin{equation}
S^x_a \sim (1 - \gamma) N^x_a
\end{equation}

Substituting this in the dynamics described by Eq.~\ref{eq:iax}:

\begin{widetext}
\begin{equation}
d_tI^x_a = -\mu_x I^x_a + \lambda_x (1 - \gamma) m N^x_a \left[ p \int da' a' \frac{I^x_{a'}}{N^x} + (1 - p) \sum_{y \ne x} \int da' a' \frac{I^y_{a'}}{N - N^y} \right]
\end{equation}
\end{widetext}

By defining $\lambda^{\text{rnd}}_x = \lambda_x (1 - \gamma)$, the equation is analogous to Eq.~\ref{eq:iax_early}. Hence, we can directly write the expression of $R_0$ in this case as: 

\begin{equation}
R^{rnd}_0 = \frac{p \sum_x \beta^{\text{rnd}}_x + \Xi^{\text{rnd}}}{\sum_x \mu_x}
\end{equation}

Where $\beta^{\text{rnd}}_x = (1 - \gamma)\beta_x$ and $\Xi^{\text{rnd}}$ is a function of $\beta^{rnd}_x$, $\mu_x$, and $c_{x,y}$.

\subsection*{Activity-based strategy}

Here, we provide the detailed derivation of the threshold for the activity-based strategy.
In this strategy, we remove all nodes of class $x$ that show an activity higher than a threshold $a_c(x)$. In practice, integrals across activities now spans from $\epsilon$ to $a_c(x)$ (and not $1$), reflecting the immunization of most active nodes. In the early stages of the spreading we can write:
\begin{equation}
    S_a^x \sim (1 - \gamma_x) N_a^x
\end{equation}
 where $\gamma_x$ is the fraction of nodes removed in class $x$. This quantity can computed as: 
\begin{equation}
\gamma_x = \int_{a_c(x)}^1 da N_a^x / N^x
\end{equation}

By repeating the same same calculation presented in Sec.~\ref{sec:sis_model} we obtain the system of $2Q$ equations:

\begin{equation}
d_t I^x = -\mu_x I^x + m \, 
\lambda_x (1 - \gamma_x) 
\left[ p \theta^x + (1 - p) \sum_{y \ne x} c_{x,y} \theta^y \right]
\end{equation}

\begin{equation}
d_t \theta^x = -\mu_x \theta^x + m \, 
\lambda_x (1 - \gamma_x) \langle a \rangle^c_x
\left[ p \theta^x + (1 - p) \sum_{y \ne x} c_{x,y} \theta^y \right]
\end{equation}

As for the previous strategy, by defining $\lambda_x^{\text{act}} = \lambda_x (1 - \gamma_x)$, we can map this system of equation to case for $\gamma=0$. Hence, we can directly write the expression for the basic reproductive number as:

\begin{equation}
R_0^{\text{act}} = \frac{p \sum_x \beta_x^{\text{act}} + \Xi^{\text{act}}}{\sum_x \mu_x}
\tag{17}
\end{equation}

Where $\beta_x^{\text{act}} = m \lambda_x^{\text{act}} \langle a \rangle^c_x$ and $\Xi^{\text{act}}$ is a function of all $\beta_x^{\text{act}}, \mu_x$, and $c_{x,y}$.

\subsection*{Egocentric sampling strategy}

Here, we provide the detailed derivation of the threshold for the egocentric sampling strategy. In this strategy, a random fraction $w$ of nodes is selected as probes. We observe their interactions (i.e., egocentric network) for $T$ time steps. Then, for each of the probes we remove, at random, exactly one of their neighbors in the aggregate egocentric network. Let us define $N_w$ as the total number of probes. Assuming a random distribution, the expected number of these in each gullibility class is $N_{w}^x = N_w \frac{N^x}{N}$. This implies that the fraction of probes in each gullibility class is $w_x=N_w^x/N_x$. In case of random distribution, the average fraction of probes in each class is the same and it is equal to the overall fraction $w_x=w$ $\forall x$.  

Let us define $P_{a}^{x}$ as the probability that, from a given probe, we select a node of activity $a$ in the gullibility class $x$. After one observation time step we can write:
\begin{eqnarray}
P_{a}^{x} &=& a p \int da' N_{a'}^x w_x \frac{m}{N_x} + \nonumber\\
&+&a (1 - p) \sum_{y\neq x}\int da' N_{a'}^{y}w_y \frac{m}{N-N_x}+ \nonumber\\
&+&\int da' a'p N_{a'}^{x}w_x\frac{m}{N_x}\frac{1}{m} \nonumber\\ 
&+&\sum_{y\neq x} \int da' a'(1 - p)N_{a'}^{y}w_y \frac{m}{N-N_y} \frac{1}{m} \nonumber\\
&=& apw_xm+a(1-p)m\frac{N_w-N_w^x}{N-N_x}+pw_x\langle a \rangle_x \nonumber\\
&+&(1-p)\sum_{y\neq x}\frac{N^y}{N-N_y} w_y \langle a \rangle_y
\end{eqnarray}

In particular, the first and the second term  represent the probability that the node is removed after reaching one of the probe, respectively, inside and outside its gullibility class; the third and the fourth term, instead, represent the probability that the node is removed after being reached from a probe, respectively, inside and outside its gullibility class. By assuming independent subsequent time steps, the probability for a node with activity $a$ and in class $x$ to be removed after $T$ periods of length $1$ is $P_a^x(T) = 1 - (1 - P_a^x)^T$. Hence, the number of nodes removed after $T$ periods with activity $a$ and in class $x$ is $R_a^x(T) = N_a^x (1 - (1 - P_a^x)^T)$. The number of susceptible in each activity and gullibility class can be approximated, at early times, $S_a^x\sim N_a^x-R_a^x$. By using these two expression in Eq.~\ref{eq:iax} integrating across all activities, we get:

\begin{equation}
d_t I^x = -\mu_x I^x + m \lambda_x \Psi_{0,x}^T 
\left[ p \theta^x + (1 - p) \sum_{y \ne x} c_{x,y} \theta^y \right] 
= h^x
\label{eq:ego1}
\end{equation}

Where we define $\Psi_{n,x}^T = \int da\, a^n\, F_x(a)(1 - P_a^x)^T$. By multiplying by $a$ and integrating across all activities instead we obtain:

\begin{equation}
d_t \theta^x = -\mu_x \theta^x + m \lambda_x \Psi_{1,x}^T 
\left[ p \theta^x + (1 - p) \sum_{y \ne x} c_{x,y} \theta^y \right] 
= g^x
\label{eq:ego2}
\end{equation}

Equations \ref{eq:ego1} and \ref{eq:ego2} define the system of $2Q$ equations in the case of egocentric network sampling immunization strategies. In this case, the mapping with the simple case with $\gamma=0$ is not immediate, at least from the system of equations. Hence, in order to obtain the threshold, we can compute the Jacobian matrix. By defining $\lambda_x^{ego} = \lambda_x \Psi_{0,x}^T$ and $\beta_x^{ego} = m \lambda_x \Psi_{1,x}^T$, we obtain:

\begin{widetext}
\[
J =
\left[
\begin{array}{cccc|ccc}
-\mu_1 & 0 & \cdots & 0 & p\lambda_1^{ego} m & (1 - p)\lambda_1^{ego} m c_{1,2} & \cdots \\
0 & -\mu_2 & \cdots & 0 & (1 - p)\lambda_2^{ego} m c_{2,1} & p \lambda_2^{ego} m & \cdots \\
\vdots & \vdots & \ddots & \vdots & \vdots & \vdots & \ddots \\
0 & 0 & \cdots & -\mu_Q & (1 - p)\lambda_Q^{ego} m c_{Q,1} & (1 - p)\lambda_Q^{ego} m c_{Q,2} & \cdots \\
\hline
0 & 0 & \cdots & 0 & -\mu_1 + p \beta_1^{ego} & (1 - p)\beta_1^{ego} c_{1,2} & \cdots \\
0 & 0 & \cdots & 0 & (1 - p)\beta_2^{ego} c_{2,1} & -\mu_2 + p \beta_2^{ego} & \cdots \\
\vdots & \vdots & \ddots & \vdots & \vdots & \vdots & \ddots \\
0 & 0 & \cdots & 0 & (1 - p)\beta_Q^{ego} c_{Q,1} & (1 - p)\beta_Q^{ego} c_{Q,2} & -\mu_Q + p \beta_Q^{ego}
\end{array}
\right]
\]
\end{widetext}

The structure of the Jacobian is very similar to the cases discussed above. Indeed, we obtain

\begin{equation}
R_0^{ego} = \frac{p \sum_x \beta_x^{ego} + \Xi^{ego}}{\sum_x \mu_x}
\end{equation}

\subsection*{Security awareness test strategy}

Here, we provide the detailed derivation of the threshold in case of the security awareness test strategy.
In this strategy, we imagine that all nodes receive a fake compromised email and/or message meant to test their awareness and susceptibility to cyber-threats. With probability $g$, a node opens the message and with probability $\lambda_x$ it gets infected, falling for the ruse. The fraction of nodes to be selected from training is picked from this subset of nodes that: 1) opened the message and 2) did not recognized it as a cyber-threat. In these settings, during the early stages of the spreading we can write $R_a^x \sim N_a^x g \lambda_x \gamma'$, where $\gamma'$ is the fraction of nodes that fall for the attack that gets selected for training. For comparability with other strategies, we now derive the expression of $\gamma'$ as function of $\gamma$. By definition of $\gamma$ we have that $\sum_x \int da \frac{R_a^x}{N} = \gamma$. By solving with respect to $\gamma'$, we obtain: 
\begin{equation}
\gamma' = \frac{\gamma}{g \langle \lambda_x \rangle},
\end{equation}
where $\langle \lambda \rangle = \sum_x \lambda_x N_x / N$ is the weighted average of the transmissibility parameter across different gullibility classes. Using this expression we get: 
\begin{equation}
R_a^x \sim N_a^x \frac{\lambda_x}{\langle \lambda_x \rangle} \gamma.
\end{equation}
Interestingly, if a class $x$ has a transmissibility parameter equal to the average, the fraction of removed nodes in that class is simply the one of the random case (i.e., $R_a^x \sim N_a^x \gamma$). If a class $x$ has transmissibility higher (lower) than the average (i.e., is more or less gullible than the average), it will have a higher (lower) fraction of removed nodes with respect to the random case. During the early stages of the spreading, the equation for $I_a^x$ can be written as:

\begin{eqnarray}
&&d_t I_{a}^{x} = -\mu_{x}I_{a}^{x} + m  \lambda_{x} N_{a}^{x} \left (1 - \frac{\lambda_x}{\langle \lambda \rangle} \gamma \right ) \times\\ \nonumber
&&
 \left [ p\int da' a' \frac{I_{a'}^{x}}{N^{x}}+(1-p)\sum_{y\neq x} \int da' a' \frac{I_{a'}^{y}}{N-N^{y}} \right ]
\end{eqnarray}

By defining $\lambda_x^{sat} = \lambda_{x}\left (1 - \frac{\lambda_x}{\langle \lambda \rangle} \gamma \right )$ we can map the equations to the simple case $\gamma=0$ for which we have already derived the solution. Hence, can directly write: 
\begin{equation}
R_{0}^{sat} = \frac{p\sum_{x}\beta_{x}^{sat} + \Xi^{sat}}{\sum_{x}\mu_{x}}, 
\end{equation}
where $\beta_{x}^{sat} = \beta_x \left (1 - \frac{\lambda_x}{\langle \lambda \rangle} \gamma \right )$ and $\Xi^{sat}$ is again a function of $\beta_{x}^{sat}$, $\mu_x$, $c_{x,y}$.

It can be shown that, for a given fraction of removed nodes $\gamma$, the average transmissibility across gullibility classes (i.e., a proxy for effectiveness of the immunization strategy) in the case of the social awareness test cannot be larger than in the random case, namely $\langle \lambda^{sat} \rangle \le \langle \lambda^{rnd} \rangle$.

As mentioned, the average transmissibility across gullibility classes is: 
\begin{align}
    \langle \lambda^{sat} \rangle &= \sum_x \lambda_x^{sat} \frac{N_x}{N} \\
    &= \sum_x \lambda_x \left(1 - \frac{\lambda_x}{\langle \lambda \rangle}\gamma \right) \frac{N_x}{N} \\
    &= \sum_x \lambda_x \frac{N_x}{N} - \sum_x \frac{\lambda_x^2}{\langle \lambda \rangle} \gamma \frac{N_x}{N} \\
    &= \langle \lambda \rangle - \gamma \frac{\langle \lambda^2 \rangle}{\langle \lambda \rangle} 
\end{align}

In the random case, instead: 
\begin{align}
\langle \lambda^{rnd} \rangle &= \sum_x \lambda_x^{rnd} \frac{N_x}{N} \\
&= \sum_x \lambda_x \left(1 -\gamma \right) \frac{N_x}{N} \\
&= \sum_x \lambda_x \frac{N_x}{N} - \gamma \sum_x \lambda_x \frac{N_x}{N} \\
&= \langle \lambda \rangle - \gamma \langle \lambda \rangle 
\end{align}
By comparing the two, we obtain that $\langle \lambda^{sat} \rangle \le \langle \lambda^{rnd} \rangle$ if the following conditions hold: 

\begin{equation}
    \langle \lambda^2 \rangle \ge \langle \lambda \rangle^2
\end{equation}

This is always verified. Even more, the equality holds only if $\lambda_x$ is constant across $x$. In other words, every time there is heterogeneity across gullibility classes, the social awareness test immunization is more efficient than the random case in reducing the spread.

\bibliographystyle{apsrev4-1}
\bibliography{main.bib}

\begin{thebibliography}{57}%
\makeatletter
\providecommand \@ifxundefined [1]{%
 \@ifx{#1\undefined}
}%
\providecommand \@ifnum [1]{%
 \ifnum #1\expandafter \@firstoftwo
 \else \expandafter \@secondoftwo
 \fi
}%
\providecommand \@ifx [1]{%
 \ifx #1\expandafter \@firstoftwo
 \else \expandafter \@secondoftwo
 \fi
}%
\providecommand \natexlab [1]{#1}%
\providecommand \enquote  [1]{``#1''}%
\providecommand \bibnamefont  [1]{#1}%
\providecommand \bibfnamefont [1]{#1}%
\providecommand \citenamefont [1]{#1}%
\providecommand \href@noop [0]{\@secondoftwo}%
\providecommand \href [0]{\begingroup \@sanitize@url \@href}%
\providecommand \@href[1]{\@@startlink{#1}\@@href}%
\providecommand \@@href[1]{\endgroup#1\@@endlink}%
\providecommand \@sanitize@url [0]{\catcode `\\12\catcode `\$12\catcode
  `\&12\catcode `\#12\catcode `\^12\catcode `\_12\catcode `\%12\relax}%
\providecommand \@@startlink[1]{}%
\providecommand \@@endlink[0]{}%
\providecommand \url  [0]{\begingroup\@sanitize@url \@url }%
\providecommand \@url [1]{\endgroup\@href {#1}{\urlprefix }}%
\providecommand \urlprefix  [0]{URL }%
\providecommand \Eprint [0]{\href }%
\providecommand \doibase [0]{http://dx.doi.org/}%
\providecommand \selectlanguage [0]{\@gobble}%
\providecommand \bibinfo  [0]{\@secondoftwo}%
\providecommand \bibfield  [0]{\@secondoftwo}%
\providecommand \translation [1]{[#1]}%
\providecommand \BibitemOpen [0]{}%
\providecommand \bibitemStop [0]{}%
\providecommand \bibitemNoStop [0]{.\EOS\space}%
\providecommand \EOS [0]{\spacefactor3000\relax}%
\providecommand \BibitemShut  [1]{\csname bibitem#1\endcsname}%
\let\auto@bib@innerbib\@empty
\bibitem [{GTR()}]{GTREPORT24}%
  \BibitemOpen
  \href@noop {} {\enquote {\bibinfo {title} {Global threat report, elastic
  security labs, 2024},}\ }\bibinfo {howpublished}
  {\url{https://www.elastic.co/explore/security-without-limits/global-threat-report}},\
  \bibinfo {note} {accessed: 2025-03-28}\BibitemShut {NoStop}%
\bibitem [{\citenamefont {Kayes}\ and\ \citenamefont
  {Iamnitchi}(2017)}]{kayes2017privacy}%
  \BibitemOpen
  \bibfield  {author} {\bibinfo {author} {\bibfnamefont {I.}~\bibnamefont
  {Kayes}}\ and\ \bibinfo {author} {\bibfnamefont {A.}~\bibnamefont
  {Iamnitchi}},\ }\href@noop {} {\bibfield  {journal} {\bibinfo  {journal}
  {Online Social Networks and Media}\ }\textbf {\bibinfo {volume} {3}},\
  \bibinfo {pages} {1} (\bibinfo {year} {2017})}\BibitemShut {NoStop}%
\bibitem [{\citenamefont {Gupta}\ \emph {et~al.}(2017)\citenamefont {Gupta},
  \citenamefont {Tewari}, \citenamefont {Jain},\ and\ \citenamefont
  {Agrawal}}]{gupta2017fighting}%
  \BibitemOpen
  \bibfield  {author} {\bibinfo {author} {\bibfnamefont {B.~B.}\ \bibnamefont
  {Gupta}}, \bibinfo {author} {\bibfnamefont {A.}~\bibnamefont {Tewari}},
  \bibinfo {author} {\bibfnamefont {A.~K.}\ \bibnamefont {Jain}}, \ and\
  \bibinfo {author} {\bibfnamefont {D.~P.}\ \bibnamefont {Agrawal}},\
  }\href@noop {} {\bibfield  {journal} {\bibinfo  {journal} {Neural Computing
  and Applications}\ }\textbf {\bibinfo {volume} {28}},\ \bibinfo {pages}
  {3629} (\bibinfo {year} {2017})}\BibitemShut {NoStop}%
\bibitem [{\citenamefont {Heartfield}\ and\ \citenamefont
  {Loukas}(2016{\natexlab{a}})}]{heartfield2016predicting}%
  \BibitemOpen
  \bibfield  {author} {\bibinfo {author} {\bibfnamefont {R.}~\bibnamefont
  {Heartfield}}\ and\ \bibinfo {author} {\bibfnamefont {G.}~\bibnamefont
  {Loukas}},\ }\href@noop {} {\bibfield  {journal} {\bibinfo  {journal}
  {International Journal on Cyber Situational Awareness (IJCSA)}\ }\textbf
  {\bibinfo {volume} {1}} (\bibinfo {year} {2016}{\natexlab{a}})}\BibitemShut
  {NoStop}%
\bibitem [{\citenamefont {Heartfield}\ \emph {et~al.}(2016)\citenamefont
  {Heartfield}, \citenamefont {Loukas},\ and\ \citenamefont
  {Gan}}]{heartfield2016you}%
  \BibitemOpen
  \bibfield  {author} {\bibinfo {author} {\bibfnamefont {R.}~\bibnamefont
  {Heartfield}}, \bibinfo {author} {\bibfnamefont {G.}~\bibnamefont {Loukas}},
  \ and\ \bibinfo {author} {\bibfnamefont {D.}~\bibnamefont {Gan}},\
  }\href@noop {} {\bibfield  {journal} {\bibinfo  {journal} {IEEE Access}\
  }\textbf {\bibinfo {volume} {4}},\ \bibinfo {pages} {6910} (\bibinfo {year}
  {2016})}\BibitemShut {NoStop}%
\bibitem [{\citenamefont {Zimmermann}\ and\ \citenamefont
  {Renaud}(2019)}]{zimmermann2019moving}%
  \BibitemOpen
  \bibfield  {author} {\bibinfo {author} {\bibfnamefont {V.}~\bibnamefont
  {Zimmermann}}\ and\ \bibinfo {author} {\bibfnamefont {K.}~\bibnamefont
  {Renaud}},\ }\href@noop {} {\bibfield  {journal} {\bibinfo  {journal}
  {International Journal of Human-Computer Studies}\ }\textbf {\bibinfo
  {volume} {131}},\ \bibinfo {pages} {169} (\bibinfo {year}
  {2019})}\BibitemShut {NoStop}%
\bibitem [{\citenamefont {Heartfield}\ and\ \citenamefont
  {Loukas}(2016{\natexlab{b}})}]{heartfield2016taxonomy}%
  \BibitemOpen
  \bibfield  {author} {\bibinfo {author} {\bibfnamefont {R.}~\bibnamefont
  {Heartfield}}\ and\ \bibinfo {author} {\bibfnamefont {G.}~\bibnamefont
  {Loukas}},\ }\href@noop {} {\bibfield  {journal} {\bibinfo  {journal} {ACM
  Computing Surveys (CSUR)}\ }\textbf {\bibinfo {volume} {48}},\ \bibinfo
  {pages} {37} (\bibinfo {year} {2016}{\natexlab{b}})}\BibitemShut {NoStop}%
\bibitem [{\citenamefont {Heartfield}\ and\ \citenamefont
  {Loukas}(2018)}]{heartfield2018protection}%
  \BibitemOpen
  \bibfield  {author} {\bibinfo {author} {\bibfnamefont {R.}~\bibnamefont
  {Heartfield}}\ and\ \bibinfo {author} {\bibfnamefont {G.}~\bibnamefont
  {Loukas}},\ }\href@noop {} {\bibfield  {journal} {\bibinfo  {journal}
  {Versatile cybersecurity}\ ,\ \bibinfo {pages} {99}} (\bibinfo {year}
  {2018})}\BibitemShut {NoStop}%
\bibitem [{\citenamefont {Falade}(2023)}]{Falade2023Decoding}%
  \BibitemOpen
  \bibfield  {author} {\bibinfo {author} {\bibfnamefont {P.}~\bibnamefont
  {Falade}},\ }\href@noop {} {\bibfield  {journal} {\bibinfo  {journal}
  {International Journal of Scientific Research in Computer Science,
  Engineering and Information Technology}\ }\textbf {\bibinfo {volume} {9}},\
  \bibinfo {pages} {5} (\bibinfo {year} {2023})}\BibitemShut {NoStop}%
\bibitem [{\citenamefont {Grbic}\ and\ \citenamefont
  {Dujlovic}(2023)}]{grbic2023social}%
  \BibitemOpen
  \bibfield  {author} {\bibinfo {author} {\bibfnamefont {D.~V.}\ \bibnamefont
  {Grbic}}\ and\ \bibinfo {author} {\bibfnamefont {I.}~\bibnamefont
  {Dujlovic}},\ }in\ \href@noop {} {\emph {\bibinfo {booktitle} {2023 22nd
  International Symposium INFOTEH-JAHORINA (INFOTEH)}}}\ (\bibinfo
  {organization} {IEEE},\ \bibinfo {year} {2023})\ pp.\ \bibinfo {pages}
  {1--5}\BibitemShut {NoStop}%
\bibitem [{\citenamefont {Holme}\ and\ \citenamefont
  {Saram{\"a}ki}(2012)}]{holme11-1}%
  \BibitemOpen
  \bibfield  {author} {\bibinfo {author} {\bibfnamefont {P.}~\bibnamefont
  {Holme}}\ and\ \bibinfo {author} {\bibfnamefont {J.}~\bibnamefont
  {Saram{\"a}ki}},\ }\href@noop {} {\bibfield  {journal} {\bibinfo  {journal}
  {Physics Reports}\ }\textbf {\bibinfo {volume} {519}},\ \bibinfo {pages} {97}
  (\bibinfo {year} {2012})}\BibitemShut {NoStop}%
\bibitem [{\citenamefont {Holme}(2015{\natexlab{a}})}]{Holme2015ModernTN}%
  \BibitemOpen
  \bibfield  {author} {\bibinfo {author} {\bibfnamefont {P.}~\bibnamefont
  {Holme}},\ }\href@noop {} {\bibfield  {journal} {\bibinfo  {journal} {The
  European Physical Journal B}\ }\textbf {\bibinfo {volume} {88}},\ \bibinfo
  {pages} {1} (\bibinfo {year} {2015}{\natexlab{a}})}\BibitemShut {NoStop}%
\bibitem [{\citenamefont {Newman}\ \emph {et~al.}(2002)\citenamefont {Newman},
  \citenamefont {Forrest},\ and\ \citenamefont {Balthrop}}]{newman2002email}%
  \BibitemOpen
  \bibfield  {author} {\bibinfo {author} {\bibfnamefont {M.~E.}\ \bibnamefont
  {Newman}}, \bibinfo {author} {\bibfnamefont {S.}~\bibnamefont {Forrest}}, \
  and\ \bibinfo {author} {\bibfnamefont {J.}~\bibnamefont {Balthrop}},\
  }\href@noop {} {\bibfield  {journal} {\bibinfo  {journal} {Physical Review
  E}\ }\textbf {\bibinfo {volume} {66}},\ \bibinfo {pages} {035101} (\bibinfo
  {year} {2002})}\BibitemShut {NoStop}%
\bibitem [{\citenamefont {Balthrop}\ \emph {et~al.}(2004)\citenamefont
  {Balthrop}, \citenamefont {Forrest}, \citenamefont {Newman},\ and\
  \citenamefont {Williamson}}]{balthrop2004technological}%
  \BibitemOpen
  \bibfield  {author} {\bibinfo {author} {\bibfnamefont {J.}~\bibnamefont
  {Balthrop}}, \bibinfo {author} {\bibfnamefont {S.}~\bibnamefont {Forrest}},
  \bibinfo {author} {\bibfnamefont {M.~E.}\ \bibnamefont {Newman}}, \ and\
  \bibinfo {author} {\bibfnamefont {M.~M.}\ \bibnamefont {Williamson}},\
  }\href@noop {} {\bibfield  {journal} {\bibinfo  {journal} {Science}\ }\textbf
  {\bibinfo {volume} {304}},\ \bibinfo {pages} {527} (\bibinfo {year}
  {2004})}\BibitemShut {NoStop}%
\bibitem [{\citenamefont {Cohen}\ \emph {et~al.}(2003)\citenamefont {Cohen},
  \citenamefont {Havlin},\ and\ \citenamefont {ben Avraham}}]{cohen03-1}%
  \BibitemOpen
  \bibfield  {author} {\bibinfo {author} {\bibfnamefont {R.}~\bibnamefont
  {Cohen}}, \bibinfo {author} {\bibfnamefont {S.}~\bibnamefont {Havlin}}, \
  and\ \bibinfo {author} {\bibfnamefont {D.}~\bibnamefont {ben Avraham}},\
  }\href@noop {} {\bibfield  {journal} {\bibinfo  {journal} {Phys Rev. Lett.}\
  }\textbf {\bibinfo {volume} {91}} (\bibinfo {year} {2003})}\BibitemShut
  {NoStop}%
\bibitem [{\citenamefont {Barrat}\ and\ \citenamefont
  {Cattuto}(2015)}]{barrat2015face}%
  \BibitemOpen
  \bibfield  {author} {\bibinfo {author} {\bibfnamefont {A.}~\bibnamefont
  {Barrat}}\ and\ \bibinfo {author} {\bibfnamefont {C.}~\bibnamefont
  {Cattuto}},\ }in\ \href@noop {} {\emph {\bibinfo {booktitle} {Social
  Phenomena}}}\ (\bibinfo  {publisher} {Springer International Publishing},\
  \bibinfo {year} {2015})\ pp.\ \bibinfo {pages} {37--57}\BibitemShut {NoStop}%
\bibitem [{\citenamefont {Perra}\ \emph
  {et~al.}(2012{\natexlab{a}})\citenamefont {Perra}, \citenamefont
  {Baronchelli}, \citenamefont {Mocanu}, \citenamefont {Gon\c{c}alves},
  \citenamefont {Pastor-Satorras},\ and\ \citenamefont
  {Vespignani}}]{perra12-2}%
  \BibitemOpen
  \bibfield  {author} {\bibinfo {author} {\bibfnamefont {N.}~\bibnamefont
  {Perra}}, \bibinfo {author} {\bibfnamefont {A.}~\bibnamefont {Baronchelli}},
  \bibinfo {author} {\bibfnamefont {D.}~\bibnamefont {Mocanu}}, \bibinfo
  {author} {\bibfnamefont {B.}~\bibnamefont {Gon\c{c}alves}}, \bibinfo {author}
  {\bibfnamefont {R.}~\bibnamefont {Pastor-Satorras}}, \ and\ \bibinfo {author}
  {\bibfnamefont {A.}~\bibnamefont {Vespignani}},\ }\href@noop {} {\bibfield
  {journal} {\bibinfo  {journal} {Physical Review Letter}\ }\textbf {\bibinfo
  {volume} {109}},\ \bibinfo {pages} {238701} (\bibinfo {year}
  {2012}{\natexlab{a}})}\BibitemShut {NoStop}%
\bibitem [{\citenamefont {Perra}\ \emph
  {et~al.}(2012{\natexlab{b}})\citenamefont {Perra}, \citenamefont
  {Gon\c{c}alves}, \citenamefont {Pastor-Satorras},\ and\ \citenamefont
  {Vespignani}}]{perra12-1}%
  \BibitemOpen
  \bibfield  {author} {\bibinfo {author} {\bibfnamefont {N.}~\bibnamefont
  {Perra}}, \bibinfo {author} {\bibfnamefont {B.}~\bibnamefont
  {Gon\c{c}alves}}, \bibinfo {author} {\bibfnamefont {R.}~\bibnamefont
  {Pastor-Satorras}}, \ and\ \bibinfo {author} {\bibfnamefont {A.}~\bibnamefont
  {Vespignani}},\ }\href@noop {} {\bibfield  {journal} {\bibinfo  {journal}
  {Scientific Reports}\ }\textbf {\bibinfo {volume} {2}},\ \bibinfo {pages}
  {469} (\bibinfo {year} {2012}{\natexlab{b}})}\BibitemShut {NoStop}%
\bibitem [{\citenamefont {Ribeiro}\ \emph {et~al.}(2013)\citenamefont
  {Ribeiro}, \citenamefont {Perra},\ and\ \citenamefont
  {Baronchelli}}]{ribeiro12-2}%
  \BibitemOpen
  \bibfield  {author} {\bibinfo {author} {\bibfnamefont {B.}~\bibnamefont
  {Ribeiro}}, \bibinfo {author} {\bibfnamefont {N.}~\bibnamefont {Perra}}, \
  and\ \bibinfo {author} {\bibfnamefont {A.}~\bibnamefont {Baronchelli}},\
  }\href@noop {} {\bibfield  {journal} {\bibinfo  {journal} {Scientific
  Reports}\ }\textbf {\bibinfo {volume} {3}},\ \bibinfo {pages} {3006}
  (\bibinfo {year} {2013})}\BibitemShut {NoStop}%
\bibitem [{\citenamefont {Liu}\ \emph {et~al.}(2014)\citenamefont {Liu},
  \citenamefont {Perra}, \citenamefont {Karsai},\ and\ \citenamefont
  {Vespignani}}]{liu2014controlling}%
  \BibitemOpen
  \bibfield  {author} {\bibinfo {author} {\bibfnamefont {S.}~\bibnamefont
  {Liu}}, \bibinfo {author} {\bibfnamefont {N.}~\bibnamefont {Perra}}, \bibinfo
  {author} {\bibfnamefont {M.}~\bibnamefont {Karsai}}, \ and\ \bibinfo {author}
  {\bibfnamefont {A.}~\bibnamefont {Vespignani}},\ }\href@noop {} {\bibfield
  {journal} {\bibinfo  {journal} {Physical review letters}\ }\textbf {\bibinfo
  {volume} {112}},\ \bibinfo {pages} {118702} (\bibinfo {year}
  {2014})}\BibitemShut {NoStop}%
\bibitem [{\citenamefont {Liu}\ \emph {et~al.}(2013)\citenamefont {Liu},
  \citenamefont {Baronchelli},\ and\ \citenamefont
  {Perra}}]{PhysRevE.87.032805}%
  \BibitemOpen
  \bibfield  {author} {\bibinfo {author} {\bibfnamefont {S.-Y.}\ \bibnamefont
  {Liu}}, \bibinfo {author} {\bibfnamefont {A.}~\bibnamefont {Baronchelli}}, \
  and\ \bibinfo {author} {\bibfnamefont {N.}~\bibnamefont {Perra}},\
  }\href@noop {} {\bibfield  {journal} {\bibinfo  {journal} {Physical Review
  E}\ }\textbf {\bibinfo {volume} {87}},\ \bibinfo {pages} {032805} (\bibinfo
  {year} {2013})}\BibitemShut {NoStop}%
\bibitem [{\citenamefont {Ren}\ and\ \citenamefont {Wang}(2014)}]{10.1063}%
  \BibitemOpen
  \bibfield  {author} {\bibinfo {author} {\bibfnamefont {G.}~\bibnamefont
  {Ren}}\ and\ \bibinfo {author} {\bibfnamefont {X.}~\bibnamefont {Wang}},\
  }\href {\doibase http://dx.doi.org/10.1063/1.4876436} {\bibfield  {journal}
  {\bibinfo  {journal} {Chaos: An Interdisciplinary Journal of Nonlinear
  Science}\ }\textbf {\bibinfo {volume} {24}},\ \bibinfo {eid} {023116}
  (\bibinfo {year} {2014})}\BibitemShut {NoStop}%
\bibitem [{\citenamefont {Starnini}\ \emph {et~al.}(2013)\citenamefont
  {Starnini}, \citenamefont {Machens}, \citenamefont {Cattuto}, \citenamefont
  {Barrat},\ and\ \citenamefont {Pastor-Satorras}}]{starnini13-1}%
  \BibitemOpen
  \bibfield  {author} {\bibinfo {author} {\bibfnamefont {M.}~\bibnamefont
  {Starnini}}, \bibinfo {author} {\bibfnamefont {A.}~\bibnamefont {Machens}},
  \bibinfo {author} {\bibfnamefont {C.}~\bibnamefont {Cattuto}}, \bibinfo
  {author} {\bibfnamefont {A.}~\bibnamefont {Barrat}}, \ and\ \bibinfo {author}
  {\bibfnamefont {R.}~\bibnamefont {Pastor-Satorras}},\ }\href@noop {}
  {\bibfield  {journal} {\bibinfo  {journal} {Journal of Theoretical Biology}\
  }\textbf {\bibinfo {volume} {337}},\ \bibinfo {pages} {89} (\bibinfo {year}
  {2013})}\BibitemShut {NoStop}%
\bibitem [{\citenamefont {Starnini}\ \emph {et~al.}(2012)\citenamefont
  {Starnini}, \citenamefont {Baronchelli}, \citenamefont {Barrat},\ and\
  \citenamefont {Pastor-Satorras}}]{starnini_rw_temp_nets}%
  \BibitemOpen
  \bibfield  {author} {\bibinfo {author} {\bibfnamefont {M.}~\bibnamefont
  {Starnini}}, \bibinfo {author} {\bibfnamefont {A.}~\bibnamefont
  {Baronchelli}}, \bibinfo {author} {\bibfnamefont {A.}~\bibnamefont {Barrat}},
  \ and\ \bibinfo {author} {\bibfnamefont {R.}~\bibnamefont
  {Pastor-Satorras}},\ }\href@noop {} {\bibfield  {journal} {\bibinfo
  {journal} {Physical Review E}\ }\textbf {\bibinfo {volume} {85}},\ \bibinfo
  {pages} {056115} (\bibinfo {year} {2012})}\BibitemShut {NoStop}%
\bibitem [{\citenamefont {Valdano}\ \emph {et~al.}(2015)\citenamefont
  {Valdano}, \citenamefont {Ferreri}, \citenamefont {Poletto},\ and\
  \citenamefont {Colizza}}]{valdano2015analytical}%
  \BibitemOpen
  \bibfield  {author} {\bibinfo {author} {\bibfnamefont {E.}~\bibnamefont
  {Valdano}}, \bibinfo {author} {\bibfnamefont {L.}~\bibnamefont {Ferreri}},
  \bibinfo {author} {\bibfnamefont {C.}~\bibnamefont {Poletto}}, \ and\
  \bibinfo {author} {\bibfnamefont {V.}~\bibnamefont {Colizza}},\ }\href@noop
  {} {\bibfield  {journal} {\bibinfo  {journal} {Physical Review X}\ }\textbf
  {\bibinfo {volume} {5}},\ \bibinfo {pages} {021005} (\bibinfo {year}
  {2015})}\BibitemShut {NoStop}%
\bibitem [{\citenamefont {Scholtes}\ \emph {et~al.}(2014)\citenamefont
  {Scholtes}, \citenamefont {Wider}, \citenamefont {Pfitzner}, \citenamefont
  {Garas}, \citenamefont {Tessone},\ and\ \citenamefont
  {Schweitzer}}]{scholtes2014causality}%
  \BibitemOpen
  \bibfield  {author} {\bibinfo {author} {\bibfnamefont {I.}~\bibnamefont
  {Scholtes}}, \bibinfo {author} {\bibfnamefont {N.}~\bibnamefont {Wider}},
  \bibinfo {author} {\bibfnamefont {R.}~\bibnamefont {Pfitzner}}, \bibinfo
  {author} {\bibfnamefont {A.}~\bibnamefont {Garas}}, \bibinfo {author}
  {\bibfnamefont {C.}~\bibnamefont {Tessone}}, \ and\ \bibinfo {author}
  {\bibfnamefont {F.}~\bibnamefont {Schweitzer}},\ }\href@noop {} {\bibfield
  {journal} {\bibinfo  {journal} {Nature Communications}\ }\textbf {\bibinfo
  {volume} {5}},\ \bibinfo {pages} {5024} (\bibinfo {year} {2014})}\BibitemShut
  {NoStop}%
\bibitem [{\citenamefont {Williams}\ and\ \citenamefont
  {Musolesi}(2016)}]{Williams160196}%
  \BibitemOpen
  \bibfield  {author} {\bibinfo {author} {\bibfnamefont {M.~J.}\ \bibnamefont
  {Williams}}\ and\ \bibinfo {author} {\bibfnamefont {M.}~\bibnamefont
  {Musolesi}},\ }\href@noop {} {\bibfield  {journal} {\bibinfo  {journal}
  {Royal Society Open Science}\ }\textbf {\bibinfo {volume} {3}},\ \bibinfo
  {pages} {160196} (\bibinfo {year} {2016})}\BibitemShut {NoStop}%
\bibitem [{\citenamefont {Rocha}\ and\ \citenamefont
  {Masuda}(2014)}]{rocha2014random}%
  \BibitemOpen
  \bibfield  {author} {\bibinfo {author} {\bibfnamefont {L.~E.}\ \bibnamefont
  {Rocha}}\ and\ \bibinfo {author} {\bibfnamefont {N.}~\bibnamefont {Masuda}},\
  }\href@noop {} {\bibfield  {journal} {\bibinfo  {journal} {New Journal of
  Physics}\ }\textbf {\bibinfo {volume} {16}},\ \bibinfo {pages} {063023}
  (\bibinfo {year} {2014})}\BibitemShut {NoStop}%
\bibitem [{\citenamefont {Takaguchi}\ \emph
  {et~al.}(2012{\natexlab{a}})\citenamefont {Takaguchi}, \citenamefont {Sato},
  \citenamefont {Yano},\ and\ \citenamefont
  {Masuda}}]{takaguchi2012importance}%
  \BibitemOpen
  \bibfield  {author} {\bibinfo {author} {\bibfnamefont {T.}~\bibnamefont
  {Takaguchi}}, \bibinfo {author} {\bibfnamefont {N.}~\bibnamefont {Sato}},
  \bibinfo {author} {\bibfnamefont {K.}~\bibnamefont {Yano}}, \ and\ \bibinfo
  {author} {\bibfnamefont {N.}~\bibnamefont {Masuda}},\ }\href@noop {}
  {\bibfield  {journal} {\bibinfo  {journal} {New Journal of Physics}\ }\textbf
  {\bibinfo {volume} {14}},\ \bibinfo {pages} {093003} (\bibinfo {year}
  {2012}{\natexlab{a}})}\BibitemShut {NoStop}%
\bibitem [{\citenamefont {Rocha}\ and\ \citenamefont
  {Blondel}(2013)}]{rocha2013bursts}%
  \BibitemOpen
  \bibfield  {author} {\bibinfo {author} {\bibfnamefont {L.~E.}\ \bibnamefont
  {Rocha}}\ and\ \bibinfo {author} {\bibfnamefont {V.~D.}\ \bibnamefont
  {Blondel}},\ }\href@noop {} {\bibfield  {journal} {\bibinfo  {journal} {PLoS
  computational biology}\ }\textbf {\bibinfo {volume} {9}},\ \bibinfo {pages}
  {e1002974} (\bibinfo {year} {2013})}\BibitemShut {NoStop}%
\bibitem [{\citenamefont {Ghoshal}\ and\ \citenamefont
  {Holme}(2006)}]{ghoshal2006attractiveness}%
  \BibitemOpen
  \bibfield  {author} {\bibinfo {author} {\bibfnamefont {G.}~\bibnamefont
  {Ghoshal}}\ and\ \bibinfo {author} {\bibfnamefont {P.}~\bibnamefont
  {Holme}},\ }\href@noop {} {\bibfield  {journal} {\bibinfo  {journal} {Physica
  A: Statistical Mechanics and its Applications}\ }\textbf {\bibinfo {volume}
  {364}},\ \bibinfo {pages} {603} (\bibinfo {year} {2006})}\BibitemShut
  {NoStop}%
\bibitem [{\citenamefont {Sun}\ \emph {et~al.}(2015)\citenamefont {Sun},
  \citenamefont {Baronchelli},\ and\ \citenamefont
  {Perra}}]{sun2015contrasting}%
  \BibitemOpen
  \bibfield  {author} {\bibinfo {author} {\bibfnamefont {K.}~\bibnamefont
  {Sun}}, \bibinfo {author} {\bibfnamefont {A.}~\bibnamefont {Baronchelli}}, \
  and\ \bibinfo {author} {\bibfnamefont {N.}~\bibnamefont {Perra}},\
  }\href@noop {} {\bibfield  {journal} {\bibinfo  {journal} {The European
  Physical Journal B}\ }\textbf {\bibinfo {volume} {88}},\ \bibinfo {pages} {1}
  (\bibinfo {year} {2015})}\BibitemShut {NoStop}%
\bibitem [{\citenamefont {Mistry}\ \emph {et~al.}(2015)\citenamefont {Mistry},
  \citenamefont {Zhang}, \citenamefont {Perra},\ and\ \citenamefont
  {Baronchelli}}]{mistry2015committed}%
  \BibitemOpen
  \bibfield  {author} {\bibinfo {author} {\bibfnamefont {D.}~\bibnamefont
  {Mistry}}, \bibinfo {author} {\bibfnamefont {Q.}~\bibnamefont {Zhang}},
  \bibinfo {author} {\bibfnamefont {N.}~\bibnamefont {Perra}}, \ and\ \bibinfo
  {author} {\bibfnamefont {A.}~\bibnamefont {Baronchelli}},\ }\href@noop {}
  {\bibfield  {journal} {\bibinfo  {journal} {Physical Review E}\ }\textbf
  {\bibinfo {volume} {92}},\ \bibinfo {pages} {042805} (\bibinfo {year}
  {2015})}\BibitemShut {NoStop}%
\bibitem [{\citenamefont {Pfitzner}\ \emph {et~al.}(2013)\citenamefont
  {Pfitzner}, \citenamefont {Scholtes}, \citenamefont {Garas}, \citenamefont
  {Tessone},\ and\ \citenamefont {Schweitzer}}]{pfitzner13-1}%
  \BibitemOpen
  \bibfield  {author} {\bibinfo {author} {\bibfnamefont {R.}~\bibnamefont
  {Pfitzner}}, \bibinfo {author} {\bibfnamefont {I.}~\bibnamefont {Scholtes}},
  \bibinfo {author} {\bibfnamefont {A.}~\bibnamefont {Garas}}, \bibinfo
  {author} {\bibfnamefont {C.}~\bibnamefont {Tessone}}, \ and\ \bibinfo
  {author} {\bibfnamefont {F.}~\bibnamefont {Schweitzer}},\ }\href@noop {}
  {\bibfield  {journal} {\bibinfo  {journal} {Physical Review Letter}\ }\textbf
  {\bibinfo {volume} {110}},\ \bibinfo {pages} {19} (\bibinfo {year}
  {2013})}\BibitemShut {NoStop}%
\bibitem [{\citenamefont {Takaguchi}\ \emph
  {et~al.}(2012{\natexlab{b}})\citenamefont {Takaguchi}, \citenamefont {Sato},
  \citenamefont {Yano},\ and\ \citenamefont {Masuda}}]{takaguchi12-1}%
  \BibitemOpen
  \bibfield  {author} {\bibinfo {author} {\bibfnamefont {T.}~\bibnamefont
  {Takaguchi}}, \bibinfo {author} {\bibfnamefont {N.}~\bibnamefont {Sato}},
  \bibinfo {author} {\bibfnamefont {K.}~\bibnamefont {Yano}}, \ and\ \bibinfo
  {author} {\bibfnamefont {N.}~\bibnamefont {Masuda}},\ }\href@noop {}
  {\bibfield  {journal} {\bibinfo  {journal} {New Journal of Physics}\ }\textbf
  {\bibinfo {volume} {14}},\ \bibinfo {pages} {093003} (\bibinfo {year}
  {2012}{\natexlab{b}})}\BibitemShut {NoStop}%
\bibitem [{\citenamefont {Takaguchi}\ \emph {et~al.}(2013)\citenamefont
  {Takaguchi}, \citenamefont {Masuda},\ and\ \citenamefont
  {Holme}}]{takaguchi2013bursty}%
  \BibitemOpen
  \bibfield  {author} {\bibinfo {author} {\bibfnamefont {T.}~\bibnamefont
  {Takaguchi}}, \bibinfo {author} {\bibfnamefont {N.}~\bibnamefont {Masuda}}, \
  and\ \bibinfo {author} {\bibfnamefont {P.}~\bibnamefont {Holme}},\
  }\href@noop {} {\bibfield  {journal} {\bibinfo  {journal} {PloS one}\
  }\textbf {\bibinfo {volume} {8}},\ \bibinfo {pages} {e68629} (\bibinfo {year}
  {2013})}\BibitemShut {NoStop}%
\bibitem [{\citenamefont {Holme}\ and\ \citenamefont
  {Liljeros}(2014)}]{holme2014birth}%
  \BibitemOpen
  \bibfield  {author} {\bibinfo {author} {\bibfnamefont {P.}~\bibnamefont
  {Holme}}\ and\ \bibinfo {author} {\bibfnamefont {F.}~\bibnamefont
  {Liljeros}},\ }\href@noop {} {\bibfield  {journal} {\bibinfo  {journal}
  {Scientific Reports}\ }\textbf {\bibinfo {volume} {4}},\ \bibinfo {pages}
  {4999} (\bibinfo {year} {2014})}\BibitemShut {NoStop}%
\bibitem [{\citenamefont {Holme}\ and\ \citenamefont
  {Masuda}(2015)}]{holme2015basic}%
  \BibitemOpen
  \bibfield  {author} {\bibinfo {author} {\bibfnamefont {P.}~\bibnamefont
  {Holme}}\ and\ \bibinfo {author} {\bibfnamefont {N.}~\bibnamefont {Masuda}},\
  }\href@noop {} {\bibfield  {journal} {\bibinfo  {journal} {PloS one}\
  }\textbf {\bibinfo {volume} {10}},\ \bibinfo {pages} {e0120567} (\bibinfo
  {year} {2015})}\BibitemShut {NoStop}%
\bibitem [{\citenamefont {Wang}\ \emph {et~al.}(2016)\citenamefont {Wang},
  \citenamefont {Bauch}, \citenamefont {Bhattacharyya}, \citenamefont
  {d'Onofrio}, \citenamefont {Manfredi}, \citenamefont {Perc}, \citenamefont
  {Perra}, \citenamefont {Salath{\'e}},\ and\ \citenamefont
  {Zhao}}]{wang2016statistical}%
  \BibitemOpen
  \bibfield  {author} {\bibinfo {author} {\bibfnamefont {Z.}~\bibnamefont
  {Wang}}, \bibinfo {author} {\bibfnamefont {C.~T.}\ \bibnamefont {Bauch}},
  \bibinfo {author} {\bibfnamefont {S.}~\bibnamefont {Bhattacharyya}}, \bibinfo
  {author} {\bibfnamefont {A.}~\bibnamefont {d'Onofrio}}, \bibinfo {author}
  {\bibfnamefont {P.}~\bibnamefont {Manfredi}}, \bibinfo {author}
  {\bibfnamefont {M.}~\bibnamefont {Perc}}, \bibinfo {author} {\bibfnamefont
  {N.}~\bibnamefont {Perra}}, \bibinfo {author} {\bibfnamefont
  {M.}~\bibnamefont {Salath{\'e}}}, \ and\ \bibinfo {author} {\bibfnamefont
  {D.}~\bibnamefont {Zhao}},\ }\href@noop {} {\bibfield  {journal} {\bibinfo
  {journal} {Physics Reports}\ }\textbf {\bibinfo {volume} {664}},\ \bibinfo
  {pages} {1} (\bibinfo {year} {2016})}\BibitemShut {NoStop}%
\bibitem [{\citenamefont {Gon{\c{c}}alves}\ and\ \citenamefont
  {Perra}(2015)}]{gonccalves2015social}%
  \BibitemOpen
  \bibfield  {author} {\bibinfo {author} {\bibfnamefont {B.}~\bibnamefont
  {Gon{\c{c}}alves}}\ and\ \bibinfo {author} {\bibfnamefont {N.}~\bibnamefont
  {Perra}},\ }\href@noop {} {\emph {\bibinfo {title} {Social phenomena: From
  data analysis to models}}}\ (\bibinfo  {publisher} {Springer},\ \bibinfo
  {year} {2015})\BibitemShut {NoStop}%
\bibitem [{\citenamefont {Brett}\ \emph {et~al.}(2019)\citenamefont {Brett},
  \citenamefont {Loukas}, \citenamefont {Moreno},\ and\ \citenamefont
  {Perra}}]{Brett_2019}%
  \BibitemOpen
  \bibfield  {author} {\bibinfo {author} {\bibfnamefont {T.}~\bibnamefont
  {Brett}}, \bibinfo {author} {\bibfnamefont {G.}~\bibnamefont {Loukas}},
  \bibinfo {author} {\bibfnamefont {Y.}~\bibnamefont {Moreno}}, \ and\ \bibinfo
  {author} {\bibfnamefont {N.}~\bibnamefont {Perra}},\ }\href@noop {}
  {\bibfield  {journal} {\bibinfo  {journal} {Physical Review E}\ }\textbf
  {\bibinfo {volume} {99}},\ \bibinfo {pages} {050303} (\bibinfo {year}
  {2019})}\BibitemShut {NoStop}%
\bibitem [{\citenamefont {Wang}\ \emph {et~al.}(2009)\citenamefont {Wang},
  \citenamefont {Gonz{\'a}lez}, \citenamefont {Hidalgo},\ and\ \citenamefont
  {Barab{\'a}si}}]{wang2009understanding}%
  \BibitemOpen
  \bibfield  {author} {\bibinfo {author} {\bibfnamefont {P.}~\bibnamefont
  {Wang}}, \bibinfo {author} {\bibfnamefont {M.~C.}\ \bibnamefont
  {Gonz{\'a}lez}}, \bibinfo {author} {\bibfnamefont {C.~A.}\ \bibnamefont
  {Hidalgo}}, \ and\ \bibinfo {author} {\bibfnamefont {A.-L.}\ \bibnamefont
  {Barab{\'a}si}},\ }\href@noop {} {\bibfield  {journal} {\bibinfo  {journal}
  {Science}\ }\textbf {\bibinfo {volume} {324}},\ \bibinfo {pages} {1071}
  (\bibinfo {year} {2009})}\BibitemShut {NoStop}%
\bibitem [{\citenamefont {Prakash}\ \emph {et~al.}(2010)\citenamefont
  {Prakash}, \citenamefont {Tong}, \citenamefont {Valler},\ and\ \citenamefont
  {Faloutsos}}]{prakash10-1}%
  \BibitemOpen
  \bibfield  {author} {\bibinfo {author} {\bibfnamefont {B.}~\bibnamefont
  {Prakash}}, \bibinfo {author} {\bibfnamefont {H.}~\bibnamefont {Tong}},
  \bibinfo {author} {\bibfnamefont {M.}~\bibnamefont {Valler}}, \ and\ \bibinfo
  {author} {\bibfnamefont {C.}~\bibnamefont {Faloutsos}},\ }\href@noop {}
  {\bibfield  {journal} {\bibinfo  {journal} {Machine Learning and Knowledge
  Discovery in Databases Lecture Notes in Computer Science}\ }\textbf {\bibinfo
  {volume} {6323}},\ \bibinfo {pages} {99} (\bibinfo {year}
  {2010})}\BibitemShut {NoStop}%
\bibitem [{\citenamefont {Heartfield}\ \emph {et~al.}(2017)\citenamefont
  {Heartfield}, \citenamefont {Loukas},\ and\ \citenamefont
  {Gan}}]{heartfield2017eye}%
  \BibitemOpen
  \bibfield  {author} {\bibinfo {author} {\bibfnamefont {R.}~\bibnamefont
  {Heartfield}}, \bibinfo {author} {\bibfnamefont {G.}~\bibnamefont {Loukas}},
  \ and\ \bibinfo {author} {\bibfnamefont {D.}~\bibnamefont {Gan}},\ }in\
  \href@noop {} {\emph {\bibinfo {booktitle} {IEEE 15th International
  Conference on Software Engineering Research, Management and Applications
  (SERA)}}}\ (\bibinfo {organization} {IEEE},\ \bibinfo {year} {2017})\ pp.\
  \bibinfo {pages} {371--378}\BibitemShut {NoStop}%
\bibitem [{\citenamefont {Peng}\ \emph {et~al.}(2017)\citenamefont {Peng},
  \citenamefont {Wang}, \citenamefont {Zhou}, \citenamefont {Wan},
  \citenamefont {Wang},\ and\ \citenamefont {Yu}}]{peng2017immunization}%
  \BibitemOpen
  \bibfield  {author} {\bibinfo {author} {\bibfnamefont {S.}~\bibnamefont
  {Peng}}, \bibinfo {author} {\bibfnamefont {G.}~\bibnamefont {Wang}}, \bibinfo
  {author} {\bibfnamefont {Y.}~\bibnamefont {Zhou}}, \bibinfo {author}
  {\bibfnamefont {C.}~\bibnamefont {Wan}}, \bibinfo {author} {\bibfnamefont
  {C.}~\bibnamefont {Wang}}, \ and\ \bibinfo {author} {\bibfnamefont
  {S.}~\bibnamefont {Yu}},\ }\href@noop {} {\bibfield  {journal} {\bibinfo
  {journal} {IEEE Transactions on Dependable and Secure Computing}\ } (\bibinfo
  {year} {2017})}\BibitemShut {NoStop}%
\bibitem [{\citenamefont {Karsai}\ \emph {et~al.}(2014)\citenamefont {Karsai},
  \citenamefont {Perra},\ and\ \citenamefont {Vespignani}}]{karsai13-1}%
  \BibitemOpen
  \bibfield  {author} {\bibinfo {author} {\bibfnamefont {M.}~\bibnamefont
  {Karsai}}, \bibinfo {author} {\bibfnamefont {N.}~\bibnamefont {Perra}}, \
  and\ \bibinfo {author} {\bibfnamefont {A.}~\bibnamefont {Vespignani}},\
  }\href@noop {} {\bibfield  {journal} {\bibinfo  {journal} {Scientific
  Reports}\ }\textbf {\bibinfo {volume} {4}},\ \bibinfo {pages} {4001}
  (\bibinfo {year} {2014})}\BibitemShut {NoStop}%
\bibitem [{\citenamefont {Ubaldi}\ \emph {et~al.}(2016)\citenamefont {Ubaldi},
  \citenamefont {Perra}, \citenamefont {Karsai}, \citenamefont {Vezzani},
  \citenamefont {Burioni},\ and\ \citenamefont
  {Vespignani}}]{ubaldi2015asymptotic}%
  \BibitemOpen
  \bibfield  {author} {\bibinfo {author} {\bibfnamefont {E.}~\bibnamefont
  {Ubaldi}}, \bibinfo {author} {\bibfnamefont {N.}~\bibnamefont {Perra}},
  \bibinfo {author} {\bibfnamefont {M.}~\bibnamefont {Karsai}}, \bibinfo
  {author} {\bibfnamefont {A.}~\bibnamefont {Vezzani}}, \bibinfo {author}
  {\bibfnamefont {R.}~\bibnamefont {Burioni}}, \ and\ \bibinfo {author}
  {\bibfnamefont {A.}~\bibnamefont {Vespignani}},\ }\href@noop {} {\bibfield
  {journal} {\bibinfo  {journal} {Scientific Reports}\ }\textbf {\bibinfo
  {volume} {6}},\ \bibinfo {pages} {35724} (\bibinfo {year}
  {2016})}\BibitemShut {NoStop}%
\bibitem [{\citenamefont {Tizzani}\ \emph {et~al.}(2018)\citenamefont
  {Tizzani}, \citenamefont {Lenti}, \citenamefont {Ubaldi}, \citenamefont
  {Vezzani}, \citenamefont {Castellano},\ and\ \citenamefont
  {Burioni}}]{tizzani2018epidemic}%
  \BibitemOpen
  \bibfield  {author} {\bibinfo {author} {\bibfnamefont {M.}~\bibnamefont
  {Tizzani}}, \bibinfo {author} {\bibfnamefont {S.}~\bibnamefont {Lenti}},
  \bibinfo {author} {\bibfnamefont {E.}~\bibnamefont {Ubaldi}}, \bibinfo
  {author} {\bibfnamefont {A.}~\bibnamefont {Vezzani}}, \bibinfo {author}
  {\bibfnamefont {C.}~\bibnamefont {Castellano}}, \ and\ \bibinfo {author}
  {\bibfnamefont {R.}~\bibnamefont {Burioni}},\ }\href@noop {} {\bibfield
  {journal} {\bibinfo  {journal} {Physical Review E}\ }\textbf {\bibinfo
  {volume} {98}},\ \bibinfo {pages} {062315} (\bibinfo {year}
  {2018})}\BibitemShut {NoStop}%
\bibitem [{\citenamefont {McPherson}\ \emph {et~al.}(2001)\citenamefont
  {McPherson}, \citenamefont {Smith-Lovin},\ and\ \citenamefont
  {Cook}}]{mcpherson2001birds}%
  \BibitemOpen
  \bibfield  {author} {\bibinfo {author} {\bibfnamefont {M.}~\bibnamefont
  {McPherson}}, \bibinfo {author} {\bibfnamefont {L.}~\bibnamefont
  {Smith-Lovin}}, \ and\ \bibinfo {author} {\bibfnamefont {J.~M.}\ \bibnamefont
  {Cook}},\ }\href@noop {} {\bibfield  {journal} {\bibinfo  {journal} {Annual
  review of sociology}\ }\textbf {\bibinfo {volume} {27}},\ \bibinfo {pages}
  {415} (\bibinfo {year} {2001})}\BibitemShut {NoStop}%
\bibitem [{\citenamefont {Keeling}\ and\ \citenamefont
  {Rohani}(2008)}]{keeling08-1}%
  \BibitemOpen
  \bibfield  {author} {\bibinfo {author} {\bibfnamefont {M.}~\bibnamefont
  {Keeling}}\ and\ \bibinfo {author} {\bibfnamefont {P.}~\bibnamefont
  {Rohani}},\ }\href@noop {} {\emph {\bibinfo {title} {Modeling Infectious
  Disease in Humans and Animals}}}\ (\bibinfo  {publisher} {Princeton
  University Press},\ \bibinfo {year} {2008})\BibitemShut {NoStop}%
\bibitem [{\citenamefont {Chernikova}\ \emph {et~al.}(2023)\citenamefont
  {Chernikova}, \citenamefont {Gozzi}, \citenamefont {Perra}, \citenamefont
  {Boboila}, \citenamefont {Eliassi-Rad},\ and\ \citenamefont
  {Oprea}}]{chernikova2023modeling}%
  \BibitemOpen
  \bibfield  {author} {\bibinfo {author} {\bibfnamefont {A.}~\bibnamefont
  {Chernikova}}, \bibinfo {author} {\bibfnamefont {N.}~\bibnamefont {Gozzi}},
  \bibinfo {author} {\bibfnamefont {N.}~\bibnamefont {Perra}}, \bibinfo
  {author} {\bibfnamefont {S.}~\bibnamefont {Boboila}}, \bibinfo {author}
  {\bibfnamefont {T.}~\bibnamefont {Eliassi-Rad}}, \ and\ \bibinfo {author}
  {\bibfnamefont {A.}~\bibnamefont {Oprea}},\ }\href@noop {} {\bibfield
  {journal} {\bibinfo  {journal} {Applied Network Science}\ }\textbf {\bibinfo
  {volume} {8}},\ \bibinfo {pages} {52} (\bibinfo {year} {2023})}\BibitemShut
  {NoStop}%
\bibitem [{\citenamefont {Newman}(2010)}]{newman10-1}%
  \BibitemOpen
  \bibfield  {author} {\bibinfo {author} {\bibfnamefont {M.}~\bibnamefont
  {Newman}},\ }\href@noop {} {\emph {\bibinfo {title} {Networks. An
  Introduction}}}\ (\bibinfo  {publisher} {Oxford Univesity Press},\ \bibinfo
  {year} {2010})\BibitemShut {NoStop}%
\bibitem [{\citenamefont {Jansson}\ and\ \citenamefont {von
  Solms}(2013)}]{jansson2013phishing}%
  \BibitemOpen
  \bibfield  {author} {\bibinfo {author} {\bibfnamefont {K.}~\bibnamefont
  {Jansson}}\ and\ \bibinfo {author} {\bibfnamefont {R.}~\bibnamefont {von
  Solms}},\ }\href@noop {} {\bibfield  {journal} {\bibinfo  {journal}
  {Behaviour \& information technology}\ }\textbf {\bibinfo {volume} {32}},\
  \bibinfo {pages} {584} (\bibinfo {year} {2013})}\BibitemShut {NoStop}%
\bibitem [{\citenamefont {Al-Daeef}\ \emph {et~al.}(2017)\citenamefont
  {Al-Daeef}, \citenamefont {Basir},\ and\ \citenamefont
  {Saudi}}]{al2017security}%
  \BibitemOpen
  \bibfield  {author} {\bibinfo {author} {\bibfnamefont {M.~M.}\ \bibnamefont
  {Al-Daeef}}, \bibinfo {author} {\bibfnamefont {N.}~\bibnamefont {Basir}}, \
  and\ \bibinfo {author} {\bibfnamefont {M.~M.}\ \bibnamefont {Saudi}},\ }in\
  \href@noop {} {\emph {\bibinfo {booktitle} {Proceedings of the world congress
  on engineering}}},\ Vol.~\bibinfo {volume} {1}\ (\bibinfo {organization}
  {WCE},\ \bibinfo {year} {2017})\ pp.\ \bibinfo {pages} {5--7}\BibitemShut
  {NoStop}%
\bibitem [{\citenamefont {Sun}\ \emph {et~al.}(2023)\citenamefont {Sun},
  \citenamefont {Ubaldi}, \citenamefont {Zhang}, \citenamefont {Karsai},\ and\
  \citenamefont {Perra}}]{sun2023effects}%
  \BibitemOpen
  \bibfield  {author} {\bibinfo {author} {\bibfnamefont {K.}~\bibnamefont
  {Sun}}, \bibinfo {author} {\bibfnamefont {E.}~\bibnamefont {Ubaldi}},
  \bibinfo {author} {\bibfnamefont {J.}~\bibnamefont {Zhang}}, \bibinfo
  {author} {\bibfnamefont {M.}~\bibnamefont {Karsai}}, \ and\ \bibinfo {author}
  {\bibfnamefont {N.}~\bibnamefont {Perra}},\ }in\ \href@noop {} {\emph
  {\bibinfo {booktitle} {Temporal Network Theory}}}\ (\bibinfo  {publisher}
  {Springer},\ \bibinfo {year} {2023})\ pp.\ \bibinfo {pages}
  {313--333}\BibitemShut {NoStop}%
\bibitem [{\citenamefont {Holme}(2015{\natexlab{b}})}]{holme2015modern}%
  \BibitemOpen
  \bibfield  {author} {\bibinfo {author} {\bibfnamefont {P.}~\bibnamefont
  {Holme}},\ }\href@noop {} {\bibfield  {journal} {\bibinfo  {journal} {The
  European Physical Journal B}\ }\textbf {\bibinfo {volume} {88}},\ \bibinfo
  {pages} {1} (\bibinfo {year} {2015}{\natexlab{b}})}\BibitemShut {NoStop}%
\bibitem [{\citenamefont {Cranford}\ \emph {et~al.}(2021)\citenamefont
  {Cranford}, \citenamefont {Gonzalez}, \citenamefont {Aggarwal}, \citenamefont
  {Tambe}, \citenamefont {Cooney},\ and\ \citenamefont
  {Lebiere}}]{cranford2021towards}%
  \BibitemOpen
  \bibfield  {author} {\bibinfo {author} {\bibfnamefont {E.~A.}\ \bibnamefont
  {Cranford}}, \bibinfo {author} {\bibfnamefont {C.}~\bibnamefont {Gonzalez}},
  \bibinfo {author} {\bibfnamefont {P.}~\bibnamefont {Aggarwal}}, \bibinfo
  {author} {\bibfnamefont {M.}~\bibnamefont {Tambe}}, \bibinfo {author}
  {\bibfnamefont {S.}~\bibnamefont {Cooney}}, \ and\ \bibinfo {author}
  {\bibfnamefont {C.}~\bibnamefont {Lebiere}},\ }\href@noop {} {\bibfield
  {journal} {\bibinfo  {journal} {Cognitive Science}\ }\textbf {\bibinfo
  {volume} {45}},\ \bibinfo {pages} {e13013} (\bibinfo {year}
  {2021})}\BibitemShut {NoStop}%
\end{thebibliography}%

\end{document}